\documentclass[12pt,preprint]{aastex}

\newcommand\simlt{\lower.5ex\hbox{$\; \buildrel < \over \sim \;$}}
\newcommand\simgt{\lower.5ex\hbox{$\; \buildrel > \over \sim \;$}}

\begin{document}

\title{Gravitational radiation from gamma-ray bursts as 
       observational opportunities for LIGO and VIRGO}
\author{Maurice H.P.M. van Putten\altaffilmark{1},
        Amir Levinson\altaffilmark{2},
        Hyun Kyu Lee\altaffilmark{3},
        Tania Regimbau\altaffilmark{1},
        Michele Punturo\altaffilmark{4},
        and Gregory M. Harry\altaffilmark{1}
}
\altaffiltext{1}{LIGO Laboratory, NW17-161, 175 Albany Street, Cambridge, MA 02139-4307, USA}
\altaffiltext{2}{School of Physics and Astronomy, Tev Aviv University, 69978 Tel Aviv, Israel, and School of
  Physics, University of Sydney, NSW2006, Australia}
\altaffiltext{3}{Department of Physics, Hanyang University 133-791, Seoul, and APCTP, Pohang, 790-784, Korea} 
\altaffiltext{4}{Virgo Project, Instituto Nazionale di Fisica Nucleare, Sezione di Perugia, Perugia, Italy}

\begin{abstract}
   Gamma-ray bursts are believed to originate 
   in core-collapse of massive stars.
   This produces an active nucleus containing a 
   rapidly rotating Kerr black hole of mass $M_H$ and angular
   velocity $\Omega_H\simeq 1/2M_H$, 
   surrounded by a uniformly magnetized torus of angular velocity 
   $\Omega_T=\eta\Omega_H$ represented 
   by two counter-oriented current rings.
   We quantify black hole-spin interactions 
   with the torus and charged particles along open magnetic 
   flux-tubes subtended by the event horizon at a finite half-opening angle 
   $\theta_H$. A major output of $E_{gw}\simeq 4\times10^{53}
   (\eta/0.1)(M_H/7M_\odot)$ erg
   is radiated in gravitational waves of frequency $f_{gw}\simeq500
   (\eta/0.1)(7M_\odot/M_H)$ Hz by a quadrupole mass-moment in the torus when
   its minor-to-major radius is less than 0.3260.
   The durations correspond to the lifetime $T_s$ of black hole-spin,
   determined by a stability condition of
   poloidal magnetic field energy-to-kinetic energy $<1/15$ in the torus.
   Consistent with GRB-SNe, we find (i) $T_s\simeq$90s (tens of s, Kouveliotou 
   et al. 1993), (ii) aspherical SNe of kinetic energy $E_{SN}\simeq2\times 10^{51}$erg
   ($2\times10^{51}$erg in SN1998bw, H\"oflich et al. 1999) and (iii) GRB-energies
   $E_\gamma\simeq2\times 10^{50}$erg ($3\times 10^{50}$erg in Frail et al. 2001),
   upon associating $\theta_H$ with poloidal curvature of the magnetosphere.
   GRB-SNe occur perhaps about once a year within $D=100$Mpc.
   Correlating LIGO/Virgo detectors enables searches for
   nearby events and their spectral closure density $6\times 10^{-9}$
   around 250Hz in the stochastic background radiation in gravitational waves. 
   At current sensitivity, LIGO-Hanford may place an upper bound around 
   150$M_\odot$ in GRB030329. Upcoming all-sky supernovae surveys may provide
   distances to GRB-SNe, conceivably coincident with weak wide-angle GRB emissions 
   similar to the nearby event GRB980425/SN1998bw. 
   Detection of $E_{gw}$ thus provides a method for 
   identifying Kerr black holes by calorimetry.
\end{abstract}


\section{Introduction}

Gravitational wave detectors LIGO \citep{abr92} and Virgo \citep{bra92},
are broad-band detectors, most sensitive in 20-1000 Hz. 
They introduce new opportunities for probing strongly-gravitating astrophysical 
sources and the stochastic background radiation in gravitational waves -- see 
\citep{cut02} for a recent overview. Notable candidates for burst sources of 
gravitational radiation are binary coalescence of neutron stars and black holes 
\citep{nar91,phi91}, newborn neutron stars \citep{fer03}, and gamma-ray bursts: 
long bursts associated with supernovae \citep{mvp01a,mes02,mvp03b} and short bursts, 
conceivably associated with black hole-neutron star coalescence. Collectively, 
these astrophysical sources may contribute appreciably to the stochastic 
background in gravitational waves. Of particular interest is the quest for
identifying Kerr black holes in the Universe, and these objects may be at the
center of cosmological GRBs.

The events GRB980425/SN1998bw \citep{gal98} and GRB030329 \citep{hjo03,sta03} 
demonstrate that long GRBs are associated with Type Ic supernovae. This provides 
considerable support for GRBs as core-collapse events in massive stars in binaries 
\citep{woo93,pac98,bro00}, and hence an association with star-forming regions. A 
correlation of the current flux-limited sample of 33 GRBs with individually measured 
redshifts with the cosmic star formation rate shows a true-to-observed GRB event rate 
of 450 \citep{mvp03c}, which is very similar to the beaming factor of 500 obtained 
from a subsample of GRBs with achromatic breaks in their light curves \citep{fra01}. 
The true GRB-event rate is hereby about one per year within a distance of 100Mpc. 
The associated supernova and late-time GRB-afterglows may present wide-angle optical 
and radio emissions as orphan transients to nearby events. Because the event rate of 
GRB980425/SN1998bw is roughly consistent with the true GRB-event rate, 
nearby events are conceivably detectable by extremely weak but non-vanishing GRB 
emissions at large viewing angles.

Here, we report on signal-to-noise ratios for gravitational radiation from
GRBs from rotating black holes in matched filtering and in correlating two
detectors of LIGO and Virgo, both in targeting GRBs as nearby point sources and in their 
contribution to the stochastic background radiation. Detection of these emissions
from GRBs with measured redshifts enables calorimetry on their inner engines, as a method for
rigorously identifying Kerr black holes as objects in the Universe.

We propose a radically new model (Fig. 1): GRB-SNe produced by an
active nucleus in a remnant stellar envelope comprising a Kerr 
black hole, surrounded by a uniformly magnetized torus in 
suspended accretion represented by two counter-oriented 
current rings \citep{mvp99,mvp01,mvp01a,mvp02b,mvp03b}. 
Our model predicts an energetic output in gravitational waves 
from black hole-spin energy with energy output and frequency
\begin{eqnarray}
E_{gw}\simeq 0.2M_\odot\left(\frac{\eta}{0.1}\right)
\left(\frac{M_H}{7M_\odot}\right),
~~~f_{gw}\simeq 500\mbox{Hz}
   \left(\frac{\eta}{0.1}\right)
   \left(\frac{7M_\odot}{M_H}\right)
\label{EQN_1}
\end{eqnarray}
emitted by a non-axisymmetric torus surrounding the black hole.
Here, we measure energy in units of $M_\odot=2\times 10^{54}$erg, and
$\eta$ denotes the ratio of the 
angular velocity of the torus to that of the black hole of mass $M_H$, which represents
the efficiency of converting black hole-spin energy into radiation by the torus. Because
the system is relativistically compact, most of the torus output is in gravitational 
radiation by multipole mass-moments \citep{mvp01a}. It has not eluded us,
that the output in gravitational radiation (\ref{EQN_1}) surpasses 
$E_\gamma\simeq 3\times 10^{50}$ erg in gamma-rays \citep{fra01}
by three orders of magnitude.

A quadrupole mass-moment associated with a mass-inhomogeneity $\delta M_T$ produces a 
luminosity \citep{pet63}
\begin{eqnarray}
L_{gw}=\frac{32}{5}\left(\omega{\cal M}\right)^{10/3}F(e)
       \simeq\frac{32}{5}\left(M_H/R\right)^5(\delta M_T/M_H)^2,
\label{EQN_PET}
\end{eqnarray}
where $\omega\simeq M^{1/2}/R^{3/2}$ denotes the orbital frequency of the torus with
major radius $R$, ${\cal M}=(\delta M_T M_H)^{3/5}/(\delta M_T+M_H)^{1/5}\simeq
M_H(\delta M_T/M_H)^{5/3}$ denotes the chirp mass, and $F(e)$ denotes a geometric
factor representing the ellipticity $e$ of the oribital motion. 
Application of (\ref{EQN_PET}) to PSR1913+16 with ellipticity $e=0.62$ \citep{hul75} 
provided the first evidence for gravitational radiation consistent with the linearized 
equations of general relativity to within 0.1\% \citep{tay94}. Here, we apply the 
right hand-side of (\ref{EQN_PET}) to a non-axisymmetric torus around a black hole,
whose mass-quadrupole inhomogeneity $\delta M_T$ is determined self-consistently in
a state of suspended accretion for the lifetime of rapid spin of the black hole. A
quadrupole mass-moment appears spontaneously as a Papaloizou-Pringle wave
\citep{pap84} whenever the torus is sufficiently slender, i.e., for a ratio
$b/R<0.3260$, where $b$ denotes the minor radius of the torus \citep{mvp02a}. 
In the suspended accretion state, most of the black hole-spin energy is dissipated 
in the event horizon for
typical ratios $\eta\sim0.1$ of the angular velocity of the torus to that of the black 
hole. Hence, the lifetime of rapid spin of the black hole is effectively determined by the
rate of dissipation of black hole-spin energy in the event horizon, itself bounded by
a finite ratio ${\cal E}_B/{\cal E}_k<1/15$ of the poloidal magnetic field energy-to-kinetic energy 
in the torus \citep{mvp03b}. This gives rise to long durations of tens of seconds for
the lifetime of rapid spin of the black hole.
The resulting gravitational wave-emissions should be limited in
band width, changing in frequency about 10\% during the emission of the first 50\% of its energy output. 
This change mirrors a decrease of 10\% in the angular velocity of a maximally-spinning 
black hole in converting 50\% of its spin-energy. Thus, gravitational radiation
is connected to Kerr black holes, representing a connection between the linearized
equations of general relativity and, respectively, fundamental objects predicted by 
the fully nonlinear equations of general relativity.

To date, short GRBs appear to have featureless afterglow emissions, and their
cosmological origin is based on an isotropic distribution in the sky and a
$<V/V_{max}>=0.385\pm0.019$ distinctly less than 1/2 \citep{kat96}. These events
are probably disconnected from star-forming regions, and may be produced by 
black hole-neutron star-coalescence \citep{pac98}, possibly associated with 
hyperaccretion onto slowly-rotating black holes \citep{mvp01}. While the wave-form of 
binary inspiral is well-understood up to 3.5 post-Newtonian order (see \cite{cut02}), 
gravitational wave-emissions from the coalescence and merger of a neutron star onto a 
slowly-rotating black hole is highly uncertain. Depending on the black hole mass
and spin, the neutron star may break up by tidal forces outside the innermost stable 
circular orbit, and subsequently form a torus which merges with the black hole. 
A torus formed from the debris of a neutron star outside the ISCO of a stellar mass 
black hole should be unstable. Quite generally, non-axisymmetries may develop as  
Papaloizou-Pringle waves \citep{pap84} in tori of finite slenderness
(see \cite{mvp03b}). In case of a massive torus formed from the debris of a
neutron star, self-gravity may also excite non-axisymmetric instabilities. 
Gravitational radiation emitted in the inspiral phase
is about 0.2$M_\odot$, followed by the emission of conceivably 0.1$M_\odot$ during the 
merger phase with the black hole. This suggests that short GRBs are potentially as 
energetic in their gravitational wave-emissions as long bursts. 

Gravitational radiation associated with collapsars has been considered in a number of
other studies \citep{nak89,moe91,bon94,dav02,fry02,min02,kob02}, also
in model-independent search strategies associated with GRBs \citep{fin99,mod02}. These studies focus on
gravitational radiation produced by the release of gravitational binding energy during
collapse and in accretion processes on a newly formed black hole (e.g, \cite{fry01}). 
We note that accretion flows are believed to be strongly turbulent, which may imply a
broad spectrum of gravitational radiation. Forementioned studies on gravitational radiation 
in core-collapse of a massive star do not invoke the spin-energy of a newly formed black hole.
They appear to indicate an energy output, which leaves a range of detectability by 
current ground based detectors of up to about 10Mpc. These events should therefore be 
considered in the context of core-collapse events independent of the GRB phenomenon, 
in light of current estimates on the local GRB event rate as referred to above. 
Currently published bounds on gravitational wave emissions from GRBs are provided
by bar detectors \citep{tri01,ast02}. These studies and results are important in identifying various
detection strategies and channels for producing gravitational waves. We suggest that the design of 
optimal detection strategies for gravitational radiation from GRBs may be facilitated by a priori 
knowledge from a specific model. 

In the presented studies, we describe a
model for GRB-SNe from rotating black holes which is consistent with 
    the observed durations and true energies in gamma-rays from 
    magnetized baryon-poor jets subtended by the event horizon of a black hole (1), 
    the observed total kinetic energies in an associated supernova, possibly 
    radio loud, with aspherical distribution of high velocity ejecta (2), and
    X-ray line-emissions produced by underlying continuum emissions (3). On this
basis, we predict band-limited gravitational wave line-emissions 
contemporaneous with the GRB according to the scaling relations (\ref{EQN_1}) at
an event rate of probably once a year within a distance of 100Mpc.

In \S2, we describe elements of current GRB-phenomenology.
In \S3-4 we summarize a theory of GRB-supernovae from rotating black holes. 
In \S5, we discuss line-broadening in response to Lense-Thirring precession.
A semi-analytical estimate is given of the contribution to the stochastic
background in gravitational waves in \S6. In \S7, we present the dimensionless
characteristic strain-amplitudes, and in \S8 we calculate the signal-to-noise ratios 
in advanced LIGO and Virgo operations in various detection strategies. \S8 introduces
a proposed detection strategy for time-frequency trajectories of slowly varying
line-emissions. We summarize our findings in \S10.

\section{Phenomenology of GRB-supernovae}

X-ray localization of GRBs by BeppoSax introduced the post-BATSE development of 
providing a sample of GRBs with individually measured redshifts (Table 1).
The recent HETE-II burst GRB 030329 has greatly enhanced 
our confidence in a GRB association to Type Ib/c supernovae, based on a similarity 
of its optical light curve and emission-lines to that of SN1998bw \citep{hjo03,sta03}. 
This observational association 
supports a GRB event rate which is locked to the star-formation rate, as core-collapse 
events in evolved massive stars \citep{woo93,pac98,bro00}. This is consistent with 
recent statistical correlations over a boad range of redshifts \citep{sch01}.
It appears that Type Ib/c SNe occur only in spiral galaxies \citep{cap97}.
Further evidence for the GRB-supernova association is found in X-ray line-emissions in 
GRB970508 \citep{piro99}, GRB970828 \citep{yos99}, GRB991216 \citep{pir00}, 
GRB000214 \citep{ant00} and GRB011211 \citep{ree02}. 

When attributing the X-ray line-emissions to excitation by
high-energy continuum emissions with energy $E_r$, \cite{ghi02}, based on \cite{laz02}, estimate 
substantial lower bounds for $E_r$: long-lived iron-line emissions in GRB99126 may require
$E_r\ge 4\times 10^{52}$erg, whilst lines from lighter elements in GRB011211 may likewise require
$E_r\ge4.4\times 10^{51}$. These lower bounds point towards an energy
reservoir in excess of that required for the true GRB-energies $E_\gamma$. We interpret this
as support for the notion that GRB inner engines may be processing other channels, which are 
contemporaneous with the baryon-poor input to the GRB-afterglow emissions.  
 
The discovery of achromatic breaks in the light curves of some of these GRBs
allowed the determination of the true GRB-energies of about $3\times 10^{50}$ 
erg, upon correcting the isotropic equivalent emissions by an observed beaming 
factor of about 500 \citep{fra01}. 
The redshift distribution of the flux-limited sample of 33 GRBs, locked to the 
SFR, allows us to estimate the unseen-to-observed GRB rate to be about 450 based
on a log-normal fit of the peak-luminosity function -- see Fig. 2 \citep{mvp03c}. 
This result is very similar to the observed beaming factor of 500. 
With the ratio 450 of the unseen-to-observed GRBs, 
the true-but-unseen GRB event rate to about $0.5\times 10^6$ per year or,
equivalently, 1 per year within a distance of 100Mpc.
This event rate agrees with that based on a fraction of about 1\% of
SNe Ib/c that might be associated with GRBs \citep{sal01}. 
Note that our event rate is a lower bound on the rate of formation of GRB inner 
engines, since we are seeing only those events in which the remnant stellar 
envelope is successfully penetrated by a baryon poor jet, e.g., when the jet
is sufficiently collimated. The true rate of formation of GRB inner engines 
a therefore an important open question.

The relatively narrow distribution of GRB-energies around $3\times 10^{50}$erg
is indicative of a standard energy reservoir \citep{fra01}. An anticorrelation
between the observed opening angle and redshift points towards
wide-angle GRB emissions which are extremely weak, as in GRB980425. Given that
the event rate of GRB980425 at $D=34$Mpc is roughly consistent with 1 per year within 
$D=100$Mpc, these wide-angle emissions may also be standard. Thus, we are led to consider 
strongly anisotropic GRB emissions in response to outflow in two directions along the
rotational axis of the progenitor star, accompanied by extremely weak GRB emissions 
in all directions (see further \cite{eic99,zha02,ros02}). In this regard,
GRB980425 $(E_{\gamma,iso}\simeq 10^{48}$erg, $z=0.0085$) is {\em not} anomalous, and GRB030329
$(E_{\gamma}\simeq 3\times10^{49}$erg, $z=0.167$) is intermediate \citep{pri03}. 
GRBs may be geometrically standard, in that this anisotropy is similar in its angular 
distribution in all sources. In this event, the inferred beaming factor depends
on redshift, i.e., is a function of the flux-limit in the sample at hand. Current 
GRB samples with individually measured redshifts, including that of Frail et al.(2001),
are dominated by sources with redshifts around unity.

Recent detection of linear polarization GRB021206 provides evidence of
synchrotron radiation in magnetized outflows, which may indicate large-scale 
magnetic fields of or produced by the inner engine \citep{cob03}. Afterglow
emissions to GRB030329 include optical emissions \citep{pri03} with 
intraday deviations from powerlaw behavior \citep{uem03}, possibly reflecting
an inhomogeneous circumburst medium or latent activity of the inner engine
\citep{che99,pri03}.

Furthermore, Type Ib/c SNe tend to be radio-loud (Turatto 2003), as in SN1990B \citep{dyk93}. 
This includes GRB980425/SN1998bw \citep{kul98,iwa99} as the
brightest Type Ib/c radio SN at a very early stage \citep{wei01}. GRB030329 
might also feature some radio emission associated with the associated SN2003dh \citep{wil03}. 
Radio emissions in these SNe are well described by optically thick (at early times) and
optically thin (at late times) synchrotron radiation of shells expanding into
a circumburst medium of stellar winds from the progenitor star \citep{li99}.
All core-collapse SNe are strongly non-spherical \citep{hoe01}, as 
in the Type II SN1987A \citep{hoe91} and in the Type Ic SN1998bw \citep{hoe99},
based, in part, on polarization measurements and direct observations. Observed
is a rotational symmetry with axis ratios of 2 to 3. This generally
reflects the presence of rotation in the progenitor star and/or in
the agent driving the explosion. Aforementioned X-ray line-emissions in GRB011211 
may be excited by high-energy continuum emissions of much larger energies \citep{ghi02}.
For Type Ib/c supernovae association with a GRB, these considerations have led some to 
suggest the presence of new explosion mechanism \citep{woo99}. 

Ultimately, GRB-supernova remnants take the form of a black hole in a binary with an optical 
companion, surrounded by a supernova remnant \citep{mvp03b}. This morphology is illustrated 
by RX J050736-6847.8 \citep{chu00}, if its X-ray binary harbors a black hole.
The accreting binary may hereby appear as a soft X-ray transient in the scenario 
of \cite{bro00}. 

\section{GRB-supernovae from rotating black holes}

GRBs are believed to be produced in core-collapse of
massive stars \citep{woo93}, whose angular momentum most likely derives from
orbital angular momentum during a common envelope stage \citep{pac98}. The 
common envelope stage must ensue only when the progenitor star is evolved 
\citep{bro00}. 
Core-collapse in this scenario describes an initial implosion which produces an 
active nucleus consisting of a Kerr black hole surrounded by a magnetized torus, 
inside a remnant stellar envelope. Evidently, the black hole-torus system is
relativistically compact, in that its linear size is on the order of its 
Schwarzschild radius. The nucleus is active, by virtue of the
spin-energy of the black hole, as outlined below.

\subsection{MeV nuclei in a remnant envelope}

We consider a uniformly magnetized torus around a rapidly rotating black hole
in its lowest energy state. This introduces radiative processes through two
spin-interactions with the black hole: a spin-connection to the torus and a 
spin-orbit coupling to charged particles along open magnetic flux-tubes.
The torus hereby catalyzes black hole-spin energy into gravitational
radiation, accompanied by winds, thermal and MeV-neutrino emissions. The open 
magnetic flux-tubes produce baryon poor outflows and subsequent high-energy 
radiation. This result is a broad range of emissions.
In what follows, $\Omega_H$ denotes the angular velocity of the
black hole and $\Omega_T$ denotes the angular velocity of the torus.

A uniformly magnetized torus introduces an ordered poloidal magnetic flux,
represented by two counter-oriented current rings. The equilibrium moment of
the black hole preserves essentially uniform and maximal horizon flux. 
When viewed in poloidal cross-section, the inner and the outer torus magnetosphere
are topologically equivalent to that of a rapidly rotating neutron star with angular velocities
$-(\Omega_H-\Omega_T)$ and $\Omega_T$, respectively. The interface between the 
inner and the outer torus magnetospheres is an ellipsoidal separatrix (Fig. 3).
An open magnetic flux-tube subtended by the event horizon of the black hole may
form by moving the separatrix of the inner and outer torus magnetosphere to
infinity \citep{mvp03b}. We emphasize that the equivalence between the inner
face of the torus and a pulsar is exact in topology, yet refers to similar but
not identical physical states. For example:
there exist corresponding annuli of $B=0$ between the last closed
field-lines of the torus and the event horizon, and between the last closed
field-lines of the pulsar and infinity, where the former features a spark gap 
which is absent in the latter; both the event horizon and asymptotic
infinity are null-surfaces, where the former but not the latter 
is endowed with finite surface gravity and the no-hair theorem.

Equivalently to pulsars, the inner face of the torus emits negative angular
momentum Alfv\'en waves into the event horizon, while the outer face emits
positive angular momentum Alfv\'en waves to infinity. Both emissions satisfy
causality. The torus hereby develops
a state of suspended accretion, described by balance of energy and angular momentum
flux received by the spin-connection to the black hole and emitted in various
channels. In response to this catalytic process, the black hole evolves by 
conservation of energy and angular momentum consistent with the no-hair 
theorem. While most of the spin-energy is dissipated in the event horizon of the 
black hole, most of the black hole-luminosity is incident onto the inner face of 
the torus. The latter represents substantial fraction of black hole-spin energy, given
by the ratio of the angular velocity of the torus to that of the black hole, and is
mostly reradiated into gravitational radiation by multipole mass moments in the 
torus. Dominant emissions in gravitational radiation are typical 
for systems whose linear size is on the order of their Schwarzshild radius, and
similar to those from new born neutron stars \citep{sha83}. The catalytic 
emissions by the torus last for the lifetime of rapid spin of the black hole.
Contemporaneously, the torus radiates a minor output in baryon-rich magnetic winds, 
thermal and MeV-neutrino emissions. The remnant stellar envelope is hereby irradiated
from {\em within} by high-energy radiation coming off the torus winds. This associated
outgoing radial momentum drives a non-spherical supernova with subsequently X-ray line-emissions when
the expanding envelope reaches optical depth of unity or less. Ultimately, this leaves 
a supernova remnant around a black hole in a binary with an optical companion. 

A spin-orbit coupling between the black hole and charged 
particles creates charged outflows along open magnetic flux-tubes subtended by the
event horizon of the black hole. Moving the separatrix in Fig. 3 to infinity by a 
stretch-fold-cut creates these open flux-tubes with a finite opening angle on the
horizon (Fig 10 in \cite{mvp03b}). In the lowest energy state of the black hole, 
the spin-orbit coupling introduces the identity \citep{haw75,mvp00}
\begin{eqnarray}
e\mbox{EMF}_\nu=\nu\Omega_H ~~\mbox{(in eV)}
\end{eqnarray}
on charged particles of angular momentum $\nu=eA_\phi$ in their Landau states along magnetic
flux-surfaces $A_\phi=$const., where $2\pi A_\phi$ denotes the magnetic flux and 
$-e$ the charge of the electron.
This coupling enables the black hole to convert mechanical work by rotation into
electrical currents {\em along} the axis of rotation. Mediated by frame-dragging,
this process is causal and local in origin, 
in that it corresponds to the line-integral of the 
electric field along the magnetic field in \cite{wal74}.
It may be compared with currents induced by Lorentz forces 
on charged particles when forced to cross magnetic field-lines, with the remarkable
distinction that the former produces an EMF parallel to and the latter produces an EMF
orthogonal to magnetic field-lines. The net current along the open flux-tube is 
determined by the detailed state of the magnetosphere formed by charge-separation and 
the boundary conditions on the horizon and at infinity, as discussed in \cite{mvp03b}.
In response to the induced charged outflows, the black hole evolves by conservation 
of energy, angular momentum and charge consistent with the no-hair theorem and,
if present, current closure. 
The fraction of black hole-spin energy released in baryon-poor outflows 
along open magnetic flux-tubes tends to be small for a finite horizon half-opening 
angle on the horizon. The outflows consist primarily of $\gamma e^\pm$, Poynting flux
and kinetic energy in baryonic contaminants. We associate these emissions with
the baryon-poor input to GRBs. These represent dissipation of kinetic energy
according to the internal shock model \citep{ree94,pir99,mes02}. Note that the
true energy
$E_\gamma\simeq 3\times 10^{50}$ \citep{fra01} in gamma-rays represents a mere 
0.01\% of the rotational energy of a stellar mass black hole.
The magnetized baryon-poor outflows are surrounded by magnetized baryon-rich
winds coming off the torus. The latter may provide collimation to the former
\citep{lev00}. Scattering of photons onto the boundary layer between the two
produces highly polarized radiation, which may exceed that attainable in
synchrotron emissions within the collimated baryon-poor jet \citep{eic03}.

Our model is parametrized as follows. The
nucleus contains a black hole of mass $M_H$, angular momentum $J_H=aM_H$ 
and electric charge $q$, where $a/M_H=\sin\lambda$ denotes the specific angular 
momentum. Because all particles approaching the event horizon assume the
angular velocity $\Omega_H=\tan(\lambda/2)/2M_H$, there is a 
magnetic moment $\mu_H=qa$ aligned with its axis of rotation \citep{car68}.
It preserves essentially uniform and maximal horizon flux at arbitrary
rotation rates in equilibrium with a surrounding torus magnetosphere of 
field-strength $B$ with $\mu_H\simeq 2BM_Hr_H^2$, where $r_H=2M_H\cos^2(\lambda/2)$ 
denotes the radius of the event horizon \citep{mvp01a}.
Upon balance with various radiation channels, the torus
develops a state of suspended accretion at MeV temperatures in equilibrium with
its input in energy and angular momentum through the spin-connection to the central black hole.
The fractions of black hole-spin energy radiated into various channels
depend on the angular velocity $\eta$ of the torus relative to that of the black 
hole, the slenderness $\delta=b/2R$ of the torus in 
terms of one-half the ratio of the minor radius $b$ to the major radius $R$, 
and the mass-fraction $\mu=M_T/M_H$ of the torus mass $M_T$ relative to $M_H$. 
The half-opening angle of the open magnetic flux-tube on the event horizon of
the black hole is denoted by $\theta_H$.

In the suspended accretion state, the torus assumes a state of differential rotation
which exceeds that of Keplerian motion. The inner face is super-Keplerian, while
the outer face is sub-Keplerian due to competing surface stresses on the inner face 
and the outer face of the torus by, respectively, the action of the black hole and 
and torus winds to infinity. Both faces may develop surface waves, similar to water
waves in channels of finite depth, since the effective gravity is outgoing in the
inner face and ingoing on the outer face. In the corotating frame, the inner and outer
faces may carry retrograde, respectively prograge waves. This allows the former
to decrease its angular momentum and the latter to increase it angular momentum. 
Any coupling between these the
inner and outer surface waves would lead to angular momentum transfer from 
the inner to the outer face, which may result in instability. 
This picture describes the Papaloizou-Pringle waves \citep{pap84},
originally discovered as an azimuthal symmetry breaking instability in tori of
infinite slenderness ($b/R\rightarrow0$). An extension of this theory to tori
of finite slenderness ($b/R=0-1$) shows that the $m\ne0$ wave-modes become
successively unstable as the torus becomes more slender. We have \citep{mvp02a}
\begin{eqnarray}
b/R<0.7506, ~0.3260,~ 0.2037,~ 0.1473,~ 0.1152,\cdots,0.56/m,
\end{eqnarray}
for the onset of instability of the $m-$th buckling mode at the point of Rayleigh stability
(stability of the $m=0$ mode between the two faces).
In the proposed suspended accretion state, the amplitude of the resulting quadrupole mass-moment,
possibly accompanied by higher order mass-moments, saturates in energy and angular momentum
balance between input from the black hole and output in forementioned radiation channels. 
These hydrodynamic instabilities may be accompanied by other instabilities, such as those associated 
with a strong magnetic field. The strength of the poloidal magnetic field-energy is subject to 
the stability criterion \citep{mvp03b}
\begin{eqnarray}
\frac{{\cal E}_B}{{\cal E}_k}<\frac{1}{15},
\label{EQN_EBK}
\end{eqnarray}
based on a linear analysis of non-axisymmetric buckling modes. A similar stability analysis
for the tilt mode replaces the right hand-side in (\ref{EQN_EBK}) with 1/12.  This
upper bound on the magnetic field-strength sets a lower bound on the dissipation rate of black hole-spin
energy in the event horizon, and hence a lower bound on the lifetime of rapid spin of the black hole.
For the parameters at hand, the lifetime of black hole-spin is hereby tens of seconds (below).
The torus itself develops MeV temperatures in a state of suspended accretion \citep{mvp03b}.

\subsection{Radiatively supernovae powered by black hole-spin energy}

The remnant stellar envelope is irradiated from within by high-energy continuum emissions from 
powerful torus winds, which were released during the preceding GRB. This continuum emission 
{\em radiatively} drives a supernova by ejection of the remnant envelope and, when the remnant
envelope has expanded sufficiently for its optical depth to this continuum emission has dropped 
below unity, excites X-ray line-emissions as observed in GRB011211 \citep{ree02,mvp03a}. 
This supernova mechanism is novel in that the supernova-energy derives ab initio from the
spin-energy of the black hole, and is otherwise similar but not identical to pulsar driven 
supernova remnants by vacuum dipole-radiation \citep{ost71}, and magnetorotational driven Type II 
supernovae by Maxwell stresses \citep{bis70,leb70,bis76,whe00,aki03} and associated heating
\citep{kun76}. 

The energy output in torus winds has been determined in a detailed calculation on the suspended
accretion state, and is found to be consistent with the lower bound of \cite{ghi02} on the energy 
in continuum emissions for the line-emissions in GRB011211 \citep{mvp03a}. In our proposed mechanism 
for supernovae with X-ray line-emissions, therefore, we envision efficient conversion of the energy output 
in torus winds into high-energy continuum emissions, possibly associated with strong shocks in the remnant 
envelope and dissipation of magnetic field-energy into radiation. We note that the latter is a long-standing 
problem in the pulsars, blazars and GRBs alike (see \cite{lev97} and references therein). Conceivably, this 
process is aided by magnetoturbulence downstream \citep{lay65,bur67}. These supernovae will be 
largely non-spherical, as determined by the collimation radius of the magnetic torus winds, see, e.g.,
\cite{cam90} and references therein.

The proposed association of the X-ray line-emissions with the supernova explosion, based on the
same underlying large energy in high-energy continuum emissions within the remnant envelope, leads
to the prediction that the intensity of line-emissions and the kinetic energy in the ejecta are
positively correlated.

\subsection{Baryon loading in the magnetized baryon-poor jet}

A small fraction of the black hole spin energy is channeled along the black hole 
rotation axis 
in the form of baryon-poor outflows along an open magnetic flux-tube, as input to the estimated 
GRB energies $E_{\gamma}=3\times 10^{50}$erg of Frail et al. (2001).  The baryon content and the loading mechanism of these jets (and essentially
of GRB fireballs in any model) is yet an open issue.  In one scenario proposed recently (Levinson
\& Eichler 2003) baryon 
loading is accomplished through pickup of neutrons diffusing into the initially baryon-free
jet from the hot, baryon-rich matter surrounding it.  The free neutrons are produced in the hot 
torus that maintains temperatures of the order of a few MeV, and stream with the baryon-rich wind
emanating from the torus to a radius of $\sim 10^{10}$ cm, above which they recombine with protons 
to form $^4$He.  The pickup process involves a collision avalanche inside the baryon-poor jet 
(BPJ), owing to
the large optical depth for inelastic nuclear collisions contributed by the inwardly diffusing 
neutrons.  The hadronic shower saturates quickly, giving rise to a viscous boundary layer at the 
outer edge of the BPJ where most of the pickups occur.  This boundary layer has a 
moderate bulk Lorentz factor.  The Lorentz factor of the BPJ core, where the baryon
density is smaller is much larger initially.  The picked-up neutrons in the hot 
boundary layer can remain free up to 
a radius of about $10^{13}$ cm where they recombine, and continue to diffuse into the BPJ core 
as the BPJ expands.  This leads to further collisions in the BPJ core with highly-relativistic 
baryons coming from 
below.  The total number of picked-up neutrons is estimated to be $\sim10^{49.5}$, although 
it depends somewhat on the outflow parameters.  The asymptotic bulk Lorentz factor of the BPJ is
established in this model at rather large radii ($\sim 10^{12}$ cm) after neutron pickup is completed, 
and lies in the range between a few hundreds to a few thousands.  The expected variation of the 
Lorentz factor across the BPJ should give rise to orientation effects that need to be assessed yet.  
The inelastic nuclear collisions 
inside the BPJ lead to efficient emission of very high-energy neutrinos (energies well above 1 TeV) 
with a very hard spectrum.  The neutrino fluxes predicted are high enough to be detected by
the upcoming km$^3$ neutrino detectors, even for a source at a redshift of 1.      
       
\section{Timescales and radiation energies}

Theoretical predictions in the model of GRBs from rotating black holes can be compared
with observations on durations and true GRB energies. We shall do so in dimensionless 
form, relative to the Newtonian timescale of orbiting matter and the rotational energy 
of a rapidly rotating Kerr black hole of mass $7M_\odot$.

The durations $T_{90}$ are given by the time of activity of the inner engine of the GRB
\citep{pir98}. We propose to identify the lifetime of the inner engine with that 
timescale $T_s$ of rapid spin of the black hole. This timescale is effectively set 
by the rate of dissipation of black hole-spin energy in the event horizon, by spin-down
against the surrounding magnetic field of strength
\begin{eqnarray}
B_c\simeq 10^{16}\mbox{G}\left(\frac{7M_\odot}{M_H}\right)\left(\frac{6M_H}{R}\right)^2
\left(\frac{M_T}{0.03M_H}\right)^{1/2}
\label{EQN_BC}
\end{eqnarray}
at the critical value in forementioned stability criterion 
${\cal E}_B/{\cal E}_k<1/15$. We note the increasing observational evidence for
super-strong magnetic fields in SGRs and AXPs, see, e.g., 
\cite{kou99,tho01,fer01,ibr01,gav02}.
We then have \citep{mvp03b}
\begin{eqnarray}
T_s\simeq 90\mbox{s} \left(\frac{M_H}{7M_\odot}\right)\left(\frac{\eta}{0.1}\right)^{-8/3}
\left(\frac{\mu}{0.03}\right)^{-1/2}.
\label{EQN_TS}
\end{eqnarray}
This estimate is consistent with durations of tens of seconds of long gamma-ray 
bursts \citep{kou93}.
This gives rise to the {\em large} parameter $\gamma_0=T_s\Omega_T$, 
\begin{eqnarray}
\gamma_0=1\times10^5
         \left(\frac{\eta}{0.1}\right)^{-8/3}
         \left(\frac{\mu}{0.03}\right)^{-1/2}
\label{EQN_G0}
\end{eqnarray}
consistent with the observed ratio $T_{90}\Omega_T\sim 10^5$.

The true energy in gamma-rays is attributed to baryon-poor energy outflow along 
an open magnetic flux tube along the axis of rotation of the black hole.
As the torus develops MeV temperatures in the suspended accretion state, it 
supports a surrounding powerful baryon-rich wind with a mass-loss rate of about
$10^{30}$g s$^{-1}$ \citep{mvp03b}. We envision that these torus winds introduce
a change in poloidal topology of the inner torus magnetosphere, upon moving
the separatrix out to infinity. This creates an open magnetic flux-tube
with finite horizon half-opening angle $\theta_H$.
The open flux-tube forms an artery for a small fraction of black hole-spin
energy, releasing magnetized baryon-poor outflows. 
For a canonical value $\epsilon\simeq 15\%$ of the efficiency of
conversion of kinetic energy-to-gamma rays (for various estimates,
see \cite{kob97,dai98,pan00,gue01}), we have, based on
\cite{mvp03b}, a {\em small} parameter
$\gamma_1=E_{\gamma}/E_{rot}$,
\begin{eqnarray}
\gamma_1\simeq \epsilon \theta_H^4.
\end{eqnarray}
Here, we propose to attribute $\theta_H$ to poloidal curvature in the inner torus 
magnetosphere, i.e., $\theta_H\simeq M_H/R$ for a magnetic field which is 
orthogonal to the polar regions of event horizon. This gives 
$E_\gamma\simeq2\times 10^{50}(\epsilon/0.15)(\eta/0.1)^{8/3}$erg, or
\begin{eqnarray}
\gamma_1 \simeq 5\times 10^{-5} \left(\frac{\epsilon}{0.15}\right)\left(\frac{\eta}{0.1}\right)^{8/3},
\end{eqnarray}
consistent with the observed ratio ${E_\gamma}/E_{rot}=7\times 10^{-5}$ for canonical 
values of $M_H=7M_\odot$ and a rapidly spinning black hole 
($E_{rot}={0.29M_H}$).

In the suspended accretion state, the torus emits correlated energies in various channels, namely 
in gravitational radiation, torus winds and thermal and MeV-neutrino emissions. 
Their fractional energies, relative to 
the rotational energy of the black hole, satisfy (\cite{mvp03a}, corrected and simplified)
\begin{eqnarray}
\gamma_2=\frac{E_{gw}}{E_{rot}}\simeq\frac{\alpha\eta}{\alpha(1+\delta)+f_w^2}\sim\eta,
\end{eqnarray}
\begin{eqnarray}
\gamma_3=\frac{E_{w}}{E_{rot}}\simeq\frac{\eta f_w^2(1-\delta)^2}{\alpha(1+\delta)+f_w^2}\sim \eta^2,
\end{eqnarray}
and, the fractional energy dissipated and converted mostly in MeV-neutrino emissions,
\begin{eqnarray}
\gamma_4=\frac{E_{diss}}{E_{rot}}\sim\delta\eta.
\label{EQN_GA}
\end{eqnarray}
The right hand-side gives the asymptotic results in the limit of strong viscosity (large $\alpha$) and small 
slenderness (small $\delta$), in case of a symmetric flux-distribution described by a 
fraction $f_w=1/2$ of open magnetic flux supported by the torus with connects to infinity. 
We remark that the strong viscosity limit satisfies $\eta\sim1/(4\alpha)$ in the limit as 
$\alpha$ becomes large. These results imply a torus temperature of about 2MeV, whereby the 
dominant emission is in MeV-neutrino emissions accompanied by subdominant thermal emissions.

The MeV nucleus is relativistically compact, whereby the dominant emission is in
gravitational radiation, rather than electromagnetic radiation. 
Its compactness can be expressed in terms of
$2\pi \int_0^{E{gw}} f_{gw}dE$, which expressed
the amount of rotational energy relative to the
linear size of the system, which is invariant under rescaling of the mass of
the black hole according the Kerr metric \citep{kerr63}. We have
(\citet{mvp01a,mvp02b}, updated with (\ref{EQN_GA}))
\begin{eqnarray}
\gamma_5 =0.0035\left(\frac{\eta}{0.1}\right)^2
\end{eqnarray}
using the trigonometric expressions $E_{rot}=2M_H\sin^2(\lambda/4)$, 
$\Omega_H=\tan(\lambda/2)/2M_H$ and $2\pi f_{gw}=2\Omega_T$, produced by spin-down 
of an extreme Kerr black hole with $\sin\lambda = a/M_H\simeq1$, where $a$ denotes 
the specific angular momentum. Values $\gamma_5>0.005$ rigorously rule out radiation 
from a rapidly-spinning neutron star, using the upper bound of $0.005$ for their 
spin-down emissions in gravitational radiation obtained from a Newtonian derivation 
for a sphere with uniform mass-density.

\subsection{A link between gravitational radiation and supernovae energies}

The gravitational wave-frequency (\ref{EQN_1}) is constrained by
the total kinetic energy in the associated supernova, and the
energy requirements for X-ray line-emissions.

The asymptotic fractional energies (\ref{EQN_GA}) introduce a relationship between the
frequency in quadrupole radiation $f_{gw}$ and the energy $E_w$ released in torus-winds, namely
\begin{eqnarray}
f_{gw}\simeq 
455\mbox{Hz}\left(\frac{E_{w}}{3.65\times 10^{52}\mbox{erg}}\right)^{1/2}
\left(\frac{7M_\odot}{M_H}\right)^{3/2},
\label{EQN_FGW}
\end{eqnarray}
where the nominal values correspond to $\eta=0.1$.
This suggests that we seek observational estimates on $E_w$ in order to constrain
the expected frequency in gravitational radiation. As mentioned in \S3, we identify
$E_w$ with the energy $E_r$ in high-energy continuum radiation which excites the 
X-ray line-emission in GRB 011211 \citep{ree02}. This points towards
\begin{eqnarray}
E_r\simeq E_w\simeq 4\times 10^{52}\mbox{erg},
\label{EQN_ER}
\end{eqnarray}
which is consistent with the required energies for $E_r$ based on \cite{ghi02}.
The kinetic energy $E_{SN}$ in the supernova ejecta is hereby identified with 
the radial momentum imparted by $E_r$ on the remnant
envelope. That is, $E_{SN}\simeq 0.5\beta E_w$, whereby
\begin{eqnarray}
E_{SN}\simeq 2\times 10^{51}\mbox{erg}\left(\frac{\beta}{0.1}\right) \left(\frac{M_H}{7M_\odot}\right)
  \left(\frac{\eta}{0.1}\right)^2
\label{EQN_SNR}
\end{eqnarray} 
with $\beta=v_{ej}/c$ denoting the velocity $v_{ej}$ of the ejecta relative to the 
velocity of light, $c$. In the expected aspherical geometry, $\beta$
refers to a mass-average of the angular distribution of the ejecta. 
The canonical value $\beta=0.1$ refers to the observed velocity of the ejecta in GRB011211. We emphasize
that $E_{SN}$ refers to the true kinetic energy in the ejecta. Eventually, the
expanding remnant envelope becomes optically thin, at which stage it may show
X-ray line-emissions excited by the underlying continuum emission $E_r$. 

In (\ref{EQN_ER}), we envision efficient conversion of the energy output 
in torus winds into high-energy continuum emissions, possibly augmented
by strong shocks in the remnant envelope and dissipation of magnetic 
field-energy into radiation. The magnetic field-strength (\ref{EQN_BC})
indicates the existence of a transition radius beyond which the magnetic
field strength becomes sub-critical. While this transition may bring about 
a change in the spectrum of radiation accompanying the torus wind, it is 
unlikely to affect conversion of wind energy to high-energy emissions at 
larger distances. The reader is referred to \cite{tho95} and \cite{dun00}
for radiative processes in super-strong magnetic fields.

We note a recent application \citep{eik03} of the suspended accretion state at keV temperatures
to explain Type B relativistic jet events in the galactic microquasar GRS 1915+105 \citep{mir94}.
We find qualitatively and quantitatively agreement with observations in energetics, timescales 
and spectral evolution, including agreement with a spectrally smooth long-duration (400-700 s) 
hard-dip state. This provides indirect support, albeit at different temperatures and densities,
for well-defined, conceivably quasi-periodic frequencies (\ref{EQN_FGW}).

\subsection{GRB980425/SN1998bw and GRB030329/SN2002dh}

In our model, {\em all} emissions are driven by the spin-energy of
the central black hole, and hence {\em all} ejecta are expected to
be non-spherical. 

The supernova explosion is non-spherical,
because the explosion energy (\ref{EQN_SNR})
represents a fraction of black hole-spin energy which is catalyzed by 
the surrounding torus (see \S3.2). $E_{SN}$ in (\ref{EQN_SNR}) is 
therefore distinct from and generally smaller than the observed isotropic 
equivalent kinetic energy $E_{k,iso}$ in the ejecta. 
Indeed, our canonical value for $E_{SN}$ agrees remarkably well with 
the estimated explosion energy of $2\times 10^{51}$ erg in 
SN1998bw \citep{hoe99}, based on asphericity in the 
anomalous expansion velocities of the ejecta. This estimate
is consistent with the partial explosion energy of about $10^{50}$ erg 
in ejecta with velocities in excess of 0.5$c$, where $c$ denotes the
velocity of light \citep{li99}. Conversely, $E_{k,iso}$ can 
readily assume anomalously large values in excess of $10^{52}$ erg,
depending on the degree of asphericity. 

In our model, the explosion energies (\ref{EQN_SNR}) 
represent normal SNe Ic values \citep{hoe99}. The term 
``hypernova" \citep{pac98} applies only to the
apparent energy $E_{k,iso}\simeq 2-3\times 10^{52}$ erg 
in GRB980425 \citep{iwa98,woo99} upon assuming 
spherical geometry, not to the true kinetic
energy $E_{SN}$ in the actual aspherical explosion.

As pointed out in \S2, the GRB emissions are strongly anisotropic,
produced by beamed baryon-poor jets along the rotational axis of 
the black hole. Based on consistency between the true GRB event rate,
based on \citep{fra01,mvp03c}, and GRB980425, we further infer 
that these beamed emissions are accompanied by extremely weak 
gamma-ray emissions over wide angles or perhaps over all directions.
The beaming factor of the baryon poor jet is about 450 
\citep{fra01,mvp03c}. Evidently, the
degree of anisotropy in the GRB emissions exceeds the axis ratio of
2 to 3 in the associated supernova ejecta \citep{hoe99} by about 
two orders of magnitude. 
While viewing the source on-axis gives rise to the brightest GRB and the largest 
$E_{k,iso}$, we conclude that viewing the source off-axis
could give rise to an apparently dim GRB with nevertheless large $E_{k,iso}$.
This may explain the apparent discrepancy between the dim
GRB980425 in the presence of a large $E_{k,iso}$, yet normal
$E_{SN}$ (\cite{hoe99}, Eqn. \ref{EQN_SNR} above), in SN1998bw.

The remarkable similarity between the optical light-curve
of SN2003dh associated with GRB030329 \citep{sta03} supports the
notion that GRBs are driven by standard inner engines. GRB030329
was a bright event in view of its proximity, though appeared with
a slightly sub-energetic $E_{\gamma, iso}$. We attribute this
to viewing strongly anisotropic GRB emissions slightly off the 
rotational axis of the black hole. Based on spectral 
data, \cite{kaw03} note that
the energy $E_{k,iso}$ of SN2003dh is probably between
that of SNe1997ef (e.g. \cite{nom01,bra01}) and SN1998bw, 
although SN2003dh and SN1998bw feature similar initial 
expansion velocities.
If SN2003dh allows a detailed aspherical model similar to that of
SN1998bw, we predict that the true kinetic energy $E_{SN}$
will attain a normal value. 

The observational constraint $E_{SNR}\simeq 2\times 10^{51}$ erg
on SN1998bw \citep{hoe99} and consistency with the energy requirement in high-energy
continuum emissions for the X-ray line-emissions in GRB011211, therefore, suggest 
an expectation value of $f_{gw}\simeq$500Hz according to (\ref{EQN_FGW}) and 
(\ref{EQN_SNR}). It would be of interest to refine this estimate by calorimetry on 
a sample of SNRs which are remnants of GRBs. Given the true GRB event of about
1 per year within a distance of 100Mpc, we anticipate about 1 GRB-SNR 
within 10Mpc. These remnants will contain a black hole in a binary with
an optical companion, possible representing a soft X-ray transient.

\section{Line-broadening from Lense-Thirring precession}

Quadrupole emissions in gravitational radiation emitted by the torus,
possibly accompanied by emissions from higher-order multipole mass-moments, 
represent a line, which changes on the secular timescale of the change in
black hole-spin. This line will broaden, when the torus precesses. 
Lense-Thirring precession \citep{len18,wil72} describes the effect of frame-dragging 
on a torus whose angular velocity vector is misaligned with the spin-axis of the black 
hole. Lense-Thirring precession is well-known in a different context, as a
possible mechanism for QPOs in X-ray binaries \citep{ste99}
as well as in black hole-neutron star binaries \citep{apo94}.
A torus which is misaligned with the spin-axis of the black hole precesses with 
essentially the frame-dragging angular velocity described by the Kerr metric. 
This is accompanied by precession of the black hole, by conservation of angular
momentum. Quite generally, the angular momentum of the torus is much less than
that of the black hole, whereby the wobbling angle of the black hole is relatively
small and can be neglected. 

Precession of the orientation of the torus modulates its cosine with the
line-of-sight. The observed strain-amplitudes are hereby phase-modulated. 
Phase-modulation of gravitational radiation from an intrinsic quadrupole moment 
introduces line-broadening. For small phase-modulations, this is manifest in
phase-coherent side-bands, which are separated from twice the orbital frequency
by the frequency of Lense-Thirring precession.
The origin of a misaligned torus may result from misaligned spin-up of the
progenitor star, prior to core-collapse, when the progenitor star is itself 
misaligned with the orbital plane of the binary.

In Boyer-Lindquist coordinates, we have to leading order the Lense-Thirring 
angular frequency $\Omega_{LT}\simeq {2J_H}/{R^3}$ for a black hole angular 
momentum $J_H=M^2\sin\lambda$ in terms of the
mass $M$ and the specific angular momentum $\sin\lambda=a/M$. Given the angular 
velocity $\Omega_H=\tan(\lambda/2)/2M$ of the black hole and
the angular velocity $\Omega_T\simeq M^{1/2}R^{-3/2}$ of the torus, we have
\begin{eqnarray}
\frac{\Omega_{LT}}{\Omega_H}\simeq 2\times 10^{-2}\left(\frac{\eta}{0.1}\right)^2
\sin^2(\lambda/2)
\label{EQN_OMH}
\end{eqnarray}
in terms of the ratio $\eta={\Omega_T}/{\Omega_H}$ of the angular velocities
of the torus to that of the black hole. We expect nominal values $\eta\sim 0.1$ 
\citep{mvp03b}, so that $\Omega_{LT}$ is about 10\% of $\Omega_T$, or, equivalently,
about 1\% of $\Omega_H$. 


An intrinsic mass-inhomogeneity $m$ in a torus introduces a luminosity
of gravitational radiation according to $L_{gw}=({32}/{5})(M/R)^5(m/M)^2,$
where ${\cal M}\simeq M(m/M)^{3/5}$ denotes the chirp mass.
The gravitational radiation thus produced is anisotropic. For each of the two
polarizations, we have
\begin{eqnarray}
h_+=\frac{4}{r}\frac{1+\cos^2\iota}{2}\cos(2\Omega_T t),~~~
h_\times=-\frac{4}{r}\cos\iota\sin(2\Omega_T t)
\label{EQN_H3}
\end{eqnarray}
where $\iota$ denotes the angle between the angular momentum and the
the line-of-sight. Precession of the torus introduces
a time-varying angle $\iota$, which modulates the strain amplitudes
$h_+$ and $h_-$ at the observer. 
Given a mean angle $\iota_0$ of the angular momentum of the torus
to the line-of-sight and a wobbling angle $\theta$, 
the time-dependent angle $\iota(t)$ of the same satisfies
\begin{eqnarray}
\cos\iota(t)=\sin\iota_0\sin\theta \cos(\omega_{LT}t) + \cos\iota_0\cos\theta.
\end{eqnarray}
Substitution of $\cos\iota(t)$ into (\ref{EQN_H3}) produces
phase-modulation. Fig. 4 illustrates the resulting
observed strain-amplitudes for various values of $\iota_0$.
We may linearize phase-modulation in the case of $\theta<<\iota_0$, 
whereby 
$h_+  =h_+^{(0)}+\theta h_+^{(1)} +O(\theta^2),$
$h_\times  =h_\times^{(0)} +\theta h_\times^{(1)} + O(\theta^2),$
where $h_+^{(0)}=h_+(\iota_0)$ and $h_\times^{(0)}=h_\times(\iota_0)$
refer to the strain-amplitudes
(\ref{EQN_H3}) with $\iota=\iota_0$,
\begin{eqnarray}
h_+^{(1)} =-\frac{\sin(2\iota_0)}{r}
\left[\cos(2\Omega_T+\Omega_{LT})+\cos(2\Omega_T-\Omega_{LT})\right]
\label{EQN_HPL}
\end{eqnarray}
and
\begin{eqnarray}
h_\times^{(1)} =\frac{2\sin(\iota_0)}{r}
\left[\sin(2\Omega_T+\Omega_{LT})+\sin(2\Omega_T-\Omega_{LT})\right].
\label{EQN_HCL}
\end{eqnarray}
The ratio of the line-strengths of the side-bands to that of the carrier 
at $2\Omega_T$ in terms of the ratio of the respective strain-amplitudes
satisfies
\begin{eqnarray}
K\simeq \theta\left(\frac{1+\cos^2\iota_0}
{1+6\cos^2\iota_0+\cos^4\iota_0}\right)^{1/2}
\sin\iota_0,
\end{eqnarray}
where we used $\Omega_{LT}<<2\Omega_{T}$. Averaged over all angles
$\iota_0$, we have $\bar{K}\simeq \theta/2$. Thus, 
a wobbling angle of about 30$^o$ typically produces
side-bands of relative strength $20\%$ (taking together
$h_+$ and $h_\times$ in each side-band). 

The above shows that Lense-Thirring precession, if present, may introduce
line-broadening by up to 5\%.

The same precession introduces time-harmonic modulation of the two
principal projections of the torus onto the celestial sphere, one
at once and one at twice the precession frequency. 
The strength of the two low-frequency lines defines the decay time
of the misalignment of the torus. These lines are extremely
small, in view of their low-frequencies, allowing Lense-Thirring to
persist for timescales at least as long as the durations of long GRBs.

\section{Stochastic background radiation from GRBs}

We may calculate the contribution of GRBs from rotating black holes to the
stochastic background in gravitational waves, for a distribution which is
locked to the star-formation rate. Below is a semi-analytic summation of the
sources, similar but not identical to the numerical summation in \citep{cow02},
and includes a correction to the amplitudes reported therein.

The spectral energy density ${d{E}_{gw}}/{df}$ of a single point source is a 
redshift-independent distribution, in view of Einstein's adiabatic relationship $E_{gw}/f$=const.
The observed energy $E_{gw}(f,z)$ at an observed frequency $f$ of a source at redshift 
$z$ hereby satisfies $E_{gw}(f,z)=(1+z)^{-1}{E_{gw}((1+z)f,0)}.$
Hence, we have $E_{gw}^\prime(f,z)=E_{gw}^\prime((1+z)f,0)$ with $^\prime=d/df$.
At redshift zero, gravitons emitted by a source at redshift $z$ are distributed over 
a surface area $4\pi d_L^2(z)$, where $d_L(z)$ denotes the
luminosity area. This gives rise to a spectral energy-density, or
equivalently, a flux per unit area at the observer, satisfying
$\tilde{F}_s(f,z)={E_{gw}^\prime}/{4\pi d_L^{2}(z)}.$
Given a star-formation rate $R_{SF}(z)$ as measured in the local rest frame
per unit of comoving volume $V$ at redshift $z$, the GRB event rate $R$ as seen by 
the observer satisfies
${dR(z)}/{dz}=\dot{n}_{GRB}(1+z)^{-1}(R_{SF}(z)/R_{SF}(0))(dV/dz),$
where $\dot{n}_{GRB}$ denotes the GRB rate-density at $z=0$.
The result contributes to the spectral energy density, i.e., flux per unit area to the 
stochastic background in gravitons by
\begin{eqnarray}
\tilde{F}_B(f)=\dot{n}_{GRB}\int_0^{z_{max}}\frac{E^\prime_{gw}\Sigma(z)}{(1+z)}dz.
\end{eqnarray}
The quantity $\Sigma(z)=(1/4\pi d_L^2(z))(R_{SF}(z)/R_{SF}(0))(dV/dz)$
is observable, representing a count-rate per unit of redshift and luminosity surface 
area (inferred with reference to a set of standard candles). Here, 
$d_L(z)=(1+z)r$ in terms of a spherical radius $r$. Considering a closed
universe of matter and vacuum energy, $\Omega_M+\Omega_\Lambda=1$, we have the
transformation rules $dV/dz=4\pi r^2/E(z,\Omega_\Lambda)$ and
$R_{SF}(z,\Omega_\Lambda)/E(z,\Omega_\Lambda)=R_{SF}(z,0)/E(z,0)$ \citep{por01}, where 
$E(z,\Omega_\Lambda)=\left[\Omega_M(1+z)^{3/2}+\Omega_\Lambda\right]^{1/2}$.
Here, we suppress a dimensionful factor $c/H_0$ in $dV/dz$,
where $c$ is the velocity of light and $H_0$ the Hubble constant. 
A cancellation in $\Sigma(z)$ leaves a model independent expression
$\Sigma(z)$ = $(1+z)^{-2}R_{SF}(z;0)R_{SF}(0;0)^{-1}E(z;0)^{-1}$, reflecting that
it is an observable. It follows that
\begin{eqnarray}
\tilde{F}_B(f)=\dot{n}_{GRB}
       \int_0^{z_{max}} \frac{{\cal F}(z;\Omega_\Lambda)}{{1+z}}
       E_{gw}^\prime((1+z)f,0)\frac{{dz}}{E(z;\Omega_\Lambda)},
\label{EQN_FB}
\end{eqnarray}
where ${\cal F}(z;\Omega_\Lambda)/(1+z)=
R_{SF}(z;\Omega_\Lambda)R_{SF}(0;\Omega_\Lambda)^{-1}(1+z)^{-3}$ 
represents an observed flux (count-rate per unit area) evolved per unit of surface area
and time at $z=0$, ab initio proportional to the star-formation rate as measured 
in the local rest frame of an Einstein-de Sitter universe per unit of comoving volume. 
The reader is referred to \citep{fer99} and \citep{phi01} for related expressions.

In the present case, we may use the comoving star-formation density \citep{mad00}
$R_{SF}(z;0)=
0.16h_{73}U(z)U(5-z)[1+660e^{-3.4(1+z)}]^{-1}M_\odot\mbox{yr}^{-1}\mbox{Mpc}^{-3}$
with Hubble constant $h_{73}$ and Heaviside function $U(\cdot)$. 
We shall further use the estimated GRB event rate of 250 yr$^{-1}$Gpc$^{-3}$, 
which includes a beaming factor of 500 \citep{fra01,mvp03c}. 

The expression (\ref{EQN_FB}) can be evaluated semi-analytically, by noting that
the gravitational wave-emissions are effectively band limited to a relative
bandwidth $B=\Delta f/f_{gw,s}$ on the order of 10\% around (\ref{EQN_1}).  
Substitution $u=1+z$ in (\ref{EQN_FB}) gives the equivalent integral
$\tilde{F}_B(f) = \dot{n}_{GRB}E_{gw}B^{-1}f_{gw,s}^{-1}
 \int_{f_{gw,s}/f(1+\xi)}^{f_{gw,s}/f(1-\xi)} D(u) du,$
where $2\xi=B$ and $D(1+z)={R_{SF}(z;\mbox{EdS})}R_{SF}^{-1}(0;\mbox{EdS})(1+z)^{-9/2}$.
To leading order, the result satisfies 
$\tilde{F}_B(f)$ $\simeq$ $\dot{n}_{GRB}$ 
$({E_{gw}}/{f}) D(f_{gw,s}/f) = \dot{n}_{GRB} \left({E_0}/{f_0}\right)  
 \left({f_0}/{f}\right)$ $\left({M}/{M_0}\right) D(f_{gw,s}/f)$
independent of $B<<1$. Here, we use the scaling relations 
\begin{eqnarray}
E_{gw}=E_0\left(\frac{M}{M_0}\right),~~~f_{gw,s}=f_0\left(\frac{M_0}{M}\right)
\label{EQN_SC2}
\end{eqnarray}
where $M_0=7M_\odot$, $E_0=0.203M_\odot(\eta/0.1)$ and $f_0=455$Hz$(\eta/0.1)$. 
Hence, we have $y=f_{gw,s}/f = f_0M_0/fM$. The average over a uniform
mass-distribution $[M_1,M_2]$ ($\Delta M=M_2-M_1$) then satisfies 
\begin{eqnarray}
<\tilde{F}_B(f)>\simeq \dot{n}_{GRB} 
  \left(\frac{E_0}{f_0}\right)\left(\frac{f_0}{f}\right) 
  \left(\frac{M_0}{\Delta M}\right) \int_{M_1}^{M_2}
  \left(\frac{f_0}{f}y^{-1}\right)D(y)\left(\frac{f_0}{f}dy^{-1}\right)
\end{eqnarray}
i.e.,
\begin{eqnarray}
<\tilde{F}_B(xf_0)> = \dot{n}_{GRB}
                     \left(\frac{E_0}{f_0}\right)
                     \left(\frac{M_0}{\Delta M}\right)f_B(x),
\label{EQN_FB0}
\end{eqnarray}
where
\begin{eqnarray}
f_B(x)=x^{-3}\int_{M_0/M_1x}^{M_0/M_0x} y^{-3}D(y)dy, ~~~x=f/f_0.
\end{eqnarray}
The function $f_B(x)=f_B(x,M_1,M_2)$ displays a broad maximum of order unity, 
reflecting the cosmological distribution $z\simeq 0-1$, preceded by a steep
rise reflecting the cosmological distribution at high redshift, and followed by
a tail $x^{-2}$ reflecting a broad distribution of mass at $z\simeq 0$ 
(Fig. 4 in \citet{cow02}). The broad maximum arises because the luminosity
factor $1/d_L^2(z)$ in energy flux effectively cancels against $dV/dz$ in this region.
In contrast, the spectral strain amplitude $\propto 1/d_L(z)$ is subdominant at low
redshift, giving rise to a sharp peak (Fig. 6 in \citet{cow02}) produced by the source 
population at intermediate redshifts $z\simeq 1$. Because $E_{gw}^\prime\propto M_H^2$, 
these peaks are dominated by high-mass sources, and, for the spectral strain amplitude, 
at about one-fourth the characteristic frequency of $f_0$.

We may average (\ref{EQN_FB0}) over a uniform mass-distribution $[M_1,M_2]=[4,14]M_\odot$, 
assuming that the black hole mass and the angular velocity ratio $\eta$ of the torus 
to that of the black are uncorrelated. Using (\ref{EQN_SC2}), we have, in dimensionful units,
\begin{eqnarray}
<\tilde{F}_B(f)>= 7.45\times10^{-9}
\hat{f}_B(x)~\frac{\mbox{erg}}{\mbox{s~}\mbox{cm}^{2}\mbox{~Hz}}
\label{EQN_FM2}
\end{eqnarray}
where $\hat{f}_B(x)=f_B(x)/\mbox{max}f_B(\cdot)$. The associated
dimensionless strain amplitude $\sqrt{S_B(f)}=(2G/\pi c^3)^{1/2}f^{-1}F_B^{1/2}(f)$,
where $G$ denotes Newton's constant, satisfies
\begin{eqnarray}
\sqrt{S_B(f)}=2.45\times 10^{-26}\left(\frac{\eta}{0.1}\right)^{-1} 
\hat{f}_S^{1/2}(x)~ \mbox{Hz}^{-1/2},
\end{eqnarray}
where $\hat{f}_S(x)=f_S(x)/\mbox{max}f_S(\cdot)$, $f_S(x)=f_B(x)/x^2$. 
Likewise, we have for the spectral closure density 
$\tilde{\Omega}_B(f)=f\tilde{F}_B(f)/\rho_cc^3$ relative to the
closure density $\rho_c=3H_0^2/8\pi G$
\begin{eqnarray}
\tilde{\Omega}_B(f)=6.11\times 10^{-9}\left(\frac{\eta}{0.1}\right)
\hat{f}_\Omega(x),
\label{EQN_OMEGA}
\end{eqnarray}
where $\hat{f}_\Omega=f_\Omega(x)/\mbox{max}f_\Omega(\cdot)$,
$f_\Omega(x)=xf_B(x)$. This shows a simple scaling relation for the 
extremal value of the spectral closure density in its dependency on
the model parameter $\eta$. The location of the maximum scales inversely 
with $f_0$, in view of $x=f/f_0$. The spectral closure density hereby 
becomes completely determined by the SFR, the fractional GRB rate thereof, 
$\eta$, and the black hole-mass distribution. Fig. 5 shows the various 
distributions. The extremal value of $\Omega_B(f)$ is in the neighborhood 
of the location of maximal sensitivity of LIGO and Virgo (see Fig. 6).

\section{Dimensionless characteristic strain amplitudes}

The strain-amplitude for a band-limited signal is commonly expressed in terms
of the dimensionless characteristic strain-amplitude of its Fourier transform.
For a signal with small relative bandwidth $B<<1$, we have (adapted from \citet{fla98})
\begin{eqnarray}
h_{\mbox{\tiny char}}=\frac{1+z}{\pi d_L(z)}\left(\frac{2E_{gw}}{f_{gw,s}B}\right)^{1/2},
\end{eqnarray}
which may be re-expressed as
\begin{eqnarray}
h_{\mbox{\tiny char}}=6.55\times 10^{-21}
\left(\frac{M}{7M_\odot}\right)\left(\frac{100\mbox{Mpc}}{d_L}\right)
\left(\frac{0.1}{B}\right)^{1/2},
\end{eqnarray}
upon ignoring dependence on redshift $z$. Note that $h_{\mbox{\tiny char}}$ is
independent of $\eta$. The signal-to-noise ratio as an expectation value over
random orientation of the source is
\begin{eqnarray}
\left(\frac{S}{N}\right)^2=\int \left(\frac{h_{\mbox{\tiny char}}}{h_n}\right)^2d\ln f
\simeq \left(\frac{h_{\mbox{\tiny char}}}{h_{\mbox{\tiny rms}}}\right)^2\frac{B}{5},
\end{eqnarray}
where $h_n=h_{rms}/\sqrt{5}$, and $h_{rms}=\sqrt{fS_h(f)}$ in terms of the spectral
noise-energy density $S_h(f)$ of the detector. The factor $1/5$ refers to 
averaging over all orientations of the source \citep{fla98}.
In light of the band-limited signal at hand, we shall consider a plot of
\begin{eqnarray}
h_{\mbox{\tiny char}}\sqrt{B/5} 
\end{eqnarray}
versus $f_{gw,s}$ according to the dependence on black hole-mass given in (\ref{EQN_SC2}),
using a canonical value $\eta=0.1$. The instantaneous spectral strain-amplitude
$h$ follows by dividing $h_{\mbox{\tiny char}}$ by the square root of the
number of $2\pi-$wave periods $N\simeq f_{gw,s}T_{90}\simeq 2\gamma_0$
according to (\ref{EQN_G0}).
It follows that
\begin{eqnarray}
h=3\times 10^{-23}
                  \left(\frac{0.1}{B}\right)^{1/2}
\left(\frac{\eta}{0.1}\right)^{4/3}\left(\frac{\mu}{0.03}\right)^{1/4}
\left(\frac{M}{7M_\odot}\right)
                  \left(\frac{100\mbox{Mpc}}{d_L}\right).
\end{eqnarray}

\section{Signal-to-noise ratios}

GRBs from rotating black holes produce emissions in the shot-noise region
of LIGO and VIRGO, where the noise strain-energy density satisfies
$S_h^{1/2}(f)\propto f$. We will discuss the signal-to-noise ratios in
various techniques. We discuss matched filtering as a theoretical upper 
bound on the achievable signal-to-noise ratios. We discuss the signal-to-noise
ratios in correlating two detectors both for searches for burst sources and
for searches for the stochastic background in gravitational radiation.

The S/N-ratio of detections using matched filtering with accurate wave-form 
templates is given by the ratio of strain amplitudes of the signal 
to that of the detector noise. Including averaging over all orientations of
the source, we have \citep{fla98,cut02}
\begin{eqnarray}
\left(\frac{S}{N}\right)_{mf}=\frac{(1+z)\sqrt{2E_{gw}}}{\pi d_L(z) f^{1/2} h_n}.
\end{eqnarray}
Here, we may neglect the redshift for distances on the order of 100Mpc. 
Consequently, for matched filtering this gives
\begin{eqnarray}
\left(\frac{S}{N}\right)_{mf} 
\simeq 8 \left(\frac{S_h^{1/2}(500\mbox{Hz})}{5.7\times 10^{-24}\mbox{Hz}^{-1/2}}\right)^{-1}
\left(\frac{\eta}{0.1}\right)^{-3/2}\left(\frac{M}{7M_\odot}\right)^{5/2}
\left(\frac{d}{100\mbox{Mpc}}\right)^{-1}.
\label{EQN_SN1}
\end{eqnarray}
The expression (\ref{EQN_SN1}) shows a strong dependence on black hole-mass. 
For a uniformly distributed mass-distribution, the we have the expectation value 
$\overline{S/N}=18$ for an average over the black hole-mass distribution 
$M_H=4-14\times M_\odot$ as observed in galactic soft X-ray transients; we have 
$\overline{S/N}=7$ for a narrower mass-distribution $M_H=5-8\times M_\odot$. The 
cumulative event rate for the resulting strain-limited sample satisfies 
$\dot{N}(S/N>s)\propto s^{-3}.$

The signal-to-noise ratio (\ref{EQN_SN1}) in matched filtering is of great
theoretical significance, in defining an upper bound in single-detector
operations. Fig. 6 shows the characteristic strain-amplitude of the 
gravitational wave-signals produced by GRBs from rotating black holes,
for a range $M=4-14\times M_\odot$ of black hole masses and a range 
$\eta=0.1-0.15$ in the ratio of the angular velocities of the torus to 
the black hole. The ratio of the characteristic strain-amplitude of a
particular event to the strain-noise amplitude of the detector (at the
same frequency) represents the signal-to-noise ratio in matched filtering.
We have included the design sensitivity curves of initial LIGO and Virgo,
and Advanced LIGO and Cryogenic Virgo. The Virgo sensitivity curve is a
current evaluation, to be validated in the coming months, during the
commissioning phase of Virgo.

Evidently, matched filtering requires detailed knowledge of the wave-form through
accurate source modeling. The magnetohydrodynamical evolution of the torus in  
the suspended accretion state has some uncertainties, such as the accompanying
accretion flow onto the torus from an extended disk. These uncertainties 
may become apparent in the gravitational wave-spectrum over long durations. 
(Similar uncertainties apply to models for gravitational radiation in accretion flows.)
For this reason, it becomes of interest to consider methods which circumvent the need
for exact wave-forms. 
In the following, we shall consider detection methods based on
the correlation of two detectors, such as the colocated pair in Hanford, or correlation
between two of the three LIGO and Virgo sites.

As mentioned in \S1, the gravitational wave-spectrum is expected to be band-limited
to within 10\% of (\ref{EQN_1}), corresponding to spin-down of a rapidly black hole 
during conversion of 50\% of its spin-energy. We may exploit this by correlating two 
detectors in narrow-band mode -- a model-independent procedure that
circumvents the need for creating wave-templates in matched filtering. An 
optimal choice of the central frequency in narrow-band mode is given by the 
expectation value of (\ref{EQN_1}) in the ensemble of GRBs from rotating black holes.

This optimal choice corresponds to the most likely value of $M_H$ and $\eta$
in our model. As indicated, present estimates indicate an optimal frequency within
0.5 to 1kHz. (A good expectation value awaits calorimetry on GRB associated supernova remnants.)
A single burst produces a spectral closure density $\Omega_s$, satisfying
$T_{90}\Omega_{s}={2E_{gw}^\prime f_{gw}}/{3H_0^2d^2}$ in geometrical units. 
The signal-to-noise ratio obtained in correlating two detector signals over an integration period $T$ 
satisfies \citep{all99}
\begin{eqnarray}
\left(\frac{S}{N}\right)^2 = \frac{9H_0^4}{50\pi^4} T
\int_0^\infty \frac{\Omega_s^2(f)df}{f^6 S_{n1}(f)S_{n2}(f)}.
\label{EQN_R}
\end{eqnarray}
This may be integrated over the bandwidth $\Delta f_{gw}<<f_{gw}$, whereby
\begin{eqnarray}
\left(\frac{S}{N}\right)\simeq \frac{1}{\sqrt{2}}\left(\frac{1}{BN}
                 \right)^{1/2}\left(\frac{S}{N}\right)^2_{mf}
\label{EQN_SNB2}
\end{eqnarray}
where $1/{BN}<1$ by the frequency-time uncertainty relation. 
The number of periods $N$ of frequency $f_{gw}$ during 
the burst of duration $T_{90}$ satisfies 
$N\simeq 2T_{90}/P\simeq 4\times 10^4\eta_{0.1}^{-8/3}\mu_{0.03}^{-1/2}.$
Hence, we have $1/BN\sim 10^{-3}$. Following (\ref{EQN_SN1}) and (\ref{EQN_R}), we find
\begin{eqnarray}
\left(\frac{S}{N}\right)\simeq 12f^{D1}_4f_4^{D2}
\left(\frac{S_h^{1/2}(500\mbox{Hz})}{5.7\times
10^{-24}\mbox{Hz}^{-1/2}}\right)^{-1}_{\mbox{D1}}
\left(\frac{S_h^{1/2}(500\mbox{Hz})}{5.7\times
10^{-24}\mbox{Hz}^{-1/2}}\right)^{-1}_{\mbox{D2}}
\eta_{0.1}^{-5/3}M_7^{5}d_8^{-2}B_{0.1}^{-1/2}\mu_{0.03}^{1/4},
\label{EQN_SN3}
\end{eqnarray}
where $\eta_{0.1}=\eta/0.1$, $M_7=M/7M_\odot$, $d_8=d/100$Mpc, 
$B_{0.1}=B/0.1$ and $\mu_{0.03}=\mu/0.03$, and the factors $f^{Di}_4=f^{Di}/4$
refer to enhancement in sensitivity in narrow-band mode, relative to 
broad-band mode. The cumulative event rate for the resulting flux-limited 
sample satisfies $\dot{N}(S/N>s)\propto s^{-3/2}.$

The S/N-ratios of (\ref{EQN_SN1}) and (\ref{EQN_SN3}) may be used to derive
upper bounds on black hole masses in GRBs, by defining a ``no-detection"
to correspond to a signal-to-noise ratio of 3 (or less). This is illustrated
in Fig. 7.

Given the proximity of the extremal value of $\Omega_B(f)$ in 
(\ref{EQN_OMEGA}) and the location of maximal sensitivity of LIGO and Virgo,
we consider correlating two colocated detectors for searches for the 
contribution of GRBs to the stochastic background in gravitational waves.
According to (\ref{EQN_R}) and (\ref{EQN_OMEGA}) for a uniform mass-distribution 
$M_H=4\times 14M_\odot$, correlation of the two advanced detectors at LIGO-Hanford
gives
\begin{eqnarray}
\left(\frac{S}{N}\right)_{B}\simeq 5 
\left(\frac{{S_h^{1/2}(500\mbox{Hz})}}{5.7\times 10^{-24}\mbox{Hz}^{-1/2}}\right)^{-1}_{\mbox{H1}}
\left(\frac{{S_h^{1/2}(500\mbox{Hz})}}{5.7\times 10^{-24}\mbox{Hz}^{-1/2}}\right)^{-1}_{\mbox{H2}}
\eta_{0.1}^{-7/2}~T_{1\mbox{yr}}^{1/2}.
\label{EQN_SNB}
\end{eqnarray}
Here, the coefficient reduces to 2.2 for a mass-distribution $M_H=5-8M_\odot$.
The estimate (\ref{EQN_SNB}) reveals an appreciable dependence on $\eta$. 

\section{A detection algorithm for time-frequency trajectories}

The proposed gravitational wave-emissions produced by GRBs from rotating black holes
are characterized by emission lines which evolve slowly in time. In this two-timing
behavior, the Newtonian timescale $T_{K}$ on the order of milliseconds serves as the 
short timescale, and the timescale $T_{s}$ of evolution of black hole-spin on the order 
of tens of seconds serves as the long timescale. In order to circumvent exact wave-form 
analysis, consider Fourier transforms on an intermediate timescale during 
which the spectrum is approximately monochromatic. Furthermore, as mentioned in the 
previous section, we may consider applying correlation techniques. In what 
follows, we consider the output of the two colocated Hanford detectors, with output
\begin{eqnarray}
s_i(t)=h(t)+n_i(t) ~~~(i=1,2),
\end{eqnarray}
where $h(t)$ denotes the strain amplitude of the source at the detector and $n_i(t)$
the strain-noise amplitude of H1 and H2.

We may take advantage of the two distinct timescales involved, by considering the
time-evolution of the spectrum evaluated over the intermediate timescale $\tau$, 
satisfying $T_K << \tau << T_s.$ We may choose $\tau$ as follows. Consider the phase 
$\Phi(t)=\omega t+(1/2)\epsilon\omega t^2$ of a line of slowly-varying frequency
$\dot{\Phi}(t)=\omega (1+ (1/2)\epsilon t)$, where $B=\epsilon T_s\simeq 0.1$ denotes
the change in frequency over the duration $T_s$ of the burst. For a duration $\tau$,
this phase evolution is essentially that of a stationary frequency, provided that 
$(1/2)\omega \epsilon \tau^2 << 2\pi,$ or
\begin{eqnarray}
{\tau}/{T_s}<<\sqrt{{2}/{BN}}\simeq 1/30.
\label{EQN_TAU}
\end{eqnarray}
For example, a typical burst duration of one minute may be divided into
$N=120$ sub-windows of 0.5s, each representing about 250 wave-periods
at a frequency of 500Hz.

Consider the discrete evolution of the spectrum of the signal of duration $T_s$ 
of the burst over $N$ sub-windows $I_n=[(n-1)\tau, n\tau]$, 
by taking successive Fourier transforms of the $s_i(t)$ over each $I_n$. 
The two spectra $\tilde{s}_i(m,n)$, where
$m$ denotes the $m-$th Fourier coefficient, can be correlated according to
\begin{eqnarray}
c(m,n)=\tilde{s}_1(m,n)\tilde{s}_2^*(m,n)+\tilde{s}_1^*(m,n)\tilde{s}_2(m,n).
\label{EQN_CMN}
\end{eqnarray}
The signal $h(t)$ contributes to a correlation between the $s_i(t)$, and hence to 
non-negative values $c_{mn}$. In general, the presence of noise introduces values
of $c_{mn}$ which are both positive and negative. Negative values of $c_{mn}$ only
appear in response to (uncorrelated) noise. A plot of positive values $c_{mn}$, therefore,
will display the evolution of the spectrum of the signal. For example, we may plot all values
of $c_{mn}$ which are greater than a certain positive number, e.g., those for which
$c_{mn}>0.3\times$max$_{mn}$$c_{mn}$. This results are illustrated in Fig. 8.

The TFT algorithm may also be applied to a single detector, i.e., LIGO at Livingston
and Virgo at Pisa, provided that the intermediate timescale (\ref{EQN_TAU}) is 
much larger than the auto-correlation time in each of the detectors. LIGO and
Virgo detectors have sample frequencies of 16kHz and 20kHz. This provides the
opportunity for down-sampling a detector signal $s(t)$ into two separate and 
interlaced sequences $s_1(t_i)$ and $s_2(t^\prime_i)$ ($t^\prime_i=t_i+\Delta t$)
that sample $f_{gw}\simeq 500$Hz, while remaining sufficiently separated for the 
noise between them to be uncorrelated. The coefficients (\ref{EQN_TAU}) would 
then be formed out of the Fourier coefficients $s_1(m,n)$ and 
$e^{im\Delta t}s_2(m,n)$.

The TFT-algorithm is of intermediate order, partly
first-order in light of the Fourier transform, and partly second-order in light of the
correlation between the Fourier coefficients of the two detector signals. Consequently,
its detection sensitivity is between matched filtering and direct correlation in the 
time-domain, as discussed in the previous section. The gain in signal-to-noise ratio
obtained in taking Fourier transforms over sub-windows may circumvent the need for
narrow-band operation, allowing use of the two Hanford detectors in their current
broad-band configuration. 

Application of the TFT algorithm to searches for the contribution of GRBs
to the stochastic background radiation could be pursued by taking the sum of the
coefficients (\ref{EQN_CMN}) over successive windows of the typical burst duration,
in light of the GRB duty cycle of about one \citep{cow02}. The contributions of the
signals from distant event add linearly, but are distributed 
over a broad range of frequencies around 250Hz. A further summation over all 
sub-windows of 0.5s would result in a net sum over $10^6$ coefficients during
a one-year observational period. The result should be an anomalous broad bumb in 
the noise around 250Hz with a signal-to-noise ratio of order unity, assuming
advanced detector sensitivity.

\section{Conclusions}

We consider signal-to-noise ratios for emissions in gravitational
radiation for GRB-SNe from rotating black holes. Our model predicts GRBs 
to be powerful burst sources for LIGO and Virgo in the frequency range above a few 
hundred Hz, described by the scaling relations (\ref{EQN_1}), and the ``big blue bar" 
in Fig. 6. Based on a true-to-observed GRB event rate ratio of about 450 
\citep{fra01,mvp03c}, the true event rate 
is estimated to be one per year within a distance of 100Mpc. Collectively, these 
events contribute a spectral energy-density $\Omega_B\simeq 6\times10^{-9}$ to the 
stochastic background in gravitational waves, which peaks around 250Hz.

Our model predicts timescales and energetics in GRBs
and provides a new mechanism for the associated supernovae (Table 2). 
The model predictions are
based on first principles and some assumptions, such as an ordered magnetic field
in the torus and a successful creation of an open, collimated magnetic flux-tube 
subtended by the event horizon of the black hole. 
The predicted durations of about 90 s (see $\gamma_0$),
radiation energies of about $2\times 10^{50}$ erg from two-sided jets (see $\gamma_1$),
and kinetic energies $2\times10^{51}$erg in non-spherical ejecta (see $\gamma_3$), 
are in excellent agreement with the observed durations of tens of seconds \citep{kou93}, 
energies $E_\gamma\simeq3\times 10^{51}$ erg in gamma-rays \citep{fra01}, and inferred 
kinetic energy $2\times 10^{51}$ erg in SN1998bw with aspherical geometry \citep{hoe99}. 
The proposed radiation driven ejection process, derived from torus wind energies,
is consistent with the energy requirement for X-ray line-emissions in GRB011211. 
This observational agreement imposes three constraints on the model
in good agreement with canonical values
$(M_H\sim 7M_\odot,\eta\sim0.1,\mu\sim0.03)$, thereby obviating the need for any
fine-tuning. We are not aware of other models for GRB inner engines which provide 
similar qualitative and quantitative agreement with a broad range of 
GRB-SNe phenomenology.

Our model predicts an output in gravitational radiation of about $4\times 10^{53}$ erg 
(see $\gamma_2$) that surpasses $E_\gamma$ by three orders of magnitude and exceeds 
the output in any other channel of emissions, including the associated SN and MeV-neutrino
emissions. This may be contrasted with Type Ia SNe, whose primary output is a few times
$10^{53}$ erg in neutrinos.
   
We have calculated the signal-to-noise ratios for the gravitational wave-emissions with
fractional energies $\gamma_2$ in various detection methods. Estimates are presented for
matched filtering, as well as narrow- and broad-band correlation techniques for nearby 
point sources. For the contribution of GRBs to the stochastic background radiation in 
gravitational waves, estimates are given for broad-band correlation between two colocated
detectors. 

For nearby point sources, matched filtering provides a theoretical upper bound in
detector sensitivity using a single-detector (which can be in broad- or narrow-band mode).
We propose to exploit the predicted narrow-band emissions by correlating two
detectors in narrow-band mode, to circumvent the need for exact wave-forms in
matched filtering. In either case, the position on the sky can be determined using 
time-of-arrival analysis at three of the LIGO and Virgo interferometers at different 
locations. Recall further that enhanced sensitivity in narrow-band
mode enhances the detection rate considerably, by increase in sensitivity volume 
beyond the loss of event rates due to frequency selection. Burst sources may also be
searched for using the proposed TFT-algorithm, whose sensitivity is 
intermediate between matched filtering and time-domain correlation techniques. This
technique takes advantage of the anticipated secular timescale in the evolution of 
the line-frequencies, and may be used in the correlation of two detectors in 
broad-band mode. The stochastic background radiation from GRBs can be searched for by
correlation between the two colocated LIGO detectors at Hanford. Direct correlation in
the time-domain gives an expected S/N=5 in broad-band mode over a one-year integration 
period. Conceivably, correlation in the Fourier domain allows an improved performance.
We note that, after instrumental line-removal, spurious correlations between the two Hanford
detectors are considered less likely in the high-frequency range than in the low
frequency range.

At current LIGO sensitivity, we may derive upper bounds to black hole-masses in
nearby GRB events, by defining a no-detection to correspond to a signal-to-noise ratio
of less than 3 in either matched filtering or correlating two detectors in broad- or 
narrow-band mode. If performed, this procedure would allow an upper bound of about $150M_\odot$ 
to be put on the mass of the black hole in GRB030229, at current LIGO Hanford sensitivity
levels in broad-band mode. We note that published bounds on strain-amplitudes from burst events 
based on bar detectors apply only to the very highest frequencies predicted by our model. 
Furthermore, the range of sensitivity implied by these limits corresponds to volume with 
negligible GRB event rates (given the estimated true GRB rate of 1 per year within 100Mpc).

The burst in gravitational radiation is expected to be contemporaneous with the GRB emissions,
extending from the time-of-onset of the GRB, or earlier on the time-scale of seconds
during which the baryon-poor outflows punched through the remnant stellar envelope
\citep{woo99}, to the end of the GRB. Nearby GRBs are conceivably observable through their
weak wide-angle emissions, similar to GRB980425 \citep{eic99}. Up to days thereafter, there 
may appear as radio supernova, representing the ejection of the remnant stellar 
envelope by the magnetic torus winds. Months thereafter, wide-angle radio 
afterglows may appear \citep{lev02,pac02}. 
Ultimately, these events leave a supernova remnant surrounding a black
hole in a binary with an optical companion \citep{mvp03b}, which may appear as 
a soft X-ray transient in the scenario of \cite{bro00}. Thus, 
long GRBs provide a unique opportunity for 
integrating LIGO and Virgo detections with current astronomical observations.

There has been a long-standing interest in gravitational wave-burst
detections coincident with GRBs \citep{fin99,mod02}.
   If GRB emissions are conical, then coincident GRB-GWB events are unlikely even
   with Advanced LIGO sensitivity, since the typical distances of observed
   GRBs are then a factor of about 8 further away than their unseen counterparts.
   In this event, we should search for events which are non-coincident with GRBs.
   However, there is increasing belief that GRB emissions are not conical. Instead,
   their emissions may be geometrically standard with strong anisotropy
   \citep{ros02,zha02}, which 
   includes extremely weak emissions extending over wide-angles 
   \citep{eic99,mvp03c}. If so, GRB980425/SN1998bw is not anomalous,
   and we may search for coincidences with such apparently weak GRBs.
At the same time, we may consider searches for the associated supernova, using
upcoming all-sky surveys such as Pan-STARRS \citep{kud03}.
The prediction of very similar time-of-onset of the burst in gravitational
radiation, weak wide-angle GRB emissions and
a radio supernova provides an important observational test for our model.

Detection of the accompanying supernova allows us to determine the distance to the 
source, and hence the energy emitted by a nearby GRB in gravitational radiation. 
This predicted energy output gives rise to a relativistic compactness parameter 
$\gamma_5\simeq2\pi E_{gw}f_{gw}$, which is predicted to be about 
$3\times 10^{-3}(\eta/0.1)^2$ in units of $c^5/G$. A sufficiently high value rigorously
rules out rapidly rotating neutron stars. 
Ultimately, detection of the proposed source of gravitational radiation provides
a method for identifying Kerr black holes in the Universe, and for 
determining their mass-range in GRBs.

{\bf Acknowledgments.} 
The authors express their thanks for constructive comments from the anonymous 
referee, the MIT-LIGO Laboratory, A. Brillet, D. Coward, R. Burman, C. Cutler, 
D. Shoemaker, R. Weiss, P. Fritchel, S. Marka, R. Araya-Gochez, L.S. Finn, R. Frey, 
P. H\"oflich and G. Tee. This work is supported by Grant No. R01-1999-00020 from  
 the Korean Science and Engineering Foundation, 
 and by the LIGO Observatories, constructed
 by Caltech and MIT with funding from NSF under cooperative agreement 
 PHY 9210038. The LIGO Laboratory operates under cooperative agreement
 PHY-0107417. This paper has been assigned LIGO document number
 LIGO-P030041-00-D.

\begin{deluxetable}{crrrrr}
\tabletypesize{\scriptsize}
\tablecaption{A sample of 33 GRBs with individually determined redshifts
\tablenotemark{a}\label{tbl-1}}
\tablewidth{0pt}
\tablehead{
GRB & Redshift $z$ & Photon flux\tablenotemark{b} & 
Luminosity \tablenotemark{c}& $\theta_j$\tablenotemark{d} & Instrument}
\startdata
970228  & 0.695 & 10      & $2.13 \times 10^{58}$ &         & SAX/WFC\\

970508  & 0.835 & 0.97    & $3.24 \times 10^{57}$ & 0.293   & SAX/WFC\\

970828  & 0.9578& 1.5     & $7.04 \times 10^{57}$ & 0.072   & RXTE/ASM\\

971214  & 3.42  & 1.96    & $2.08 \times 10^{59}$ & $>0.056$& SAX/WFC\\

980425  & 0.0085& 0.96    & $1.54 \times 10^{53}$ &         & SAX/WFC\\

980613  & 1.096 & 0.5     & $3.28 \times 10^{57}$ & $>0.127$& SAX/WFC\\

980703  & 0.966 & 2.40    & $1.15 \times 10^{58}$ & 0.135   & RXTE/ASM\\

990123  & 1.6   & 16.41   & $2.74 \times 10^{59}$ & 0.050   & SAX/WFC\\

990506  & 1.3   & 18.56   & $1.85 \times 10^{59}$ &         & BAT/PCA\\

990510  & 1.619 & 8.16    & $1.40 \times 10^{59}$ & 0.053   & SAX/WFC\\

990705  & 0.86  &         &                     & 0.054   & SAX/WFC\\

990712  & 0.434 & 11.64   & $7.97 \times 10^{57}$ & $>0.411$& SAX/WFC\\

991208  & 0.706 & 11.2*   & $2.48 \times 10^{58}$ & $<0.079$& Uly/KO/NE\\

991216  & 1.02  & 67.5    & $3.70 \times 10^{59}$ & 0.051   & BAT/PCA\\

000131  & 4.5   & 1.5*    & $3.05 \times 10^{59}$ &$<0.047$ & Uly/KO/NE\\

000210  & 0.846 & 29.9    & $1.03 \times 10^{59}$ &         & SAX/WFC\\

000301C & 0.42  & 1.32*   & $8.37 \times 10^{56}$ & 0.105   & ASM/Uly\\

000214  & 2.03  &         &                     &         & SAX/WFC\\

000418  & 1.118 & 3.3*    & $2.27 \times 10^{58}$ & 0.198   & Uly/KO/NE \\

000911  & 1.058 & 2.86    & $1.72 \times 10^{58}$ &         & Uly/KO/NE\\

000926  & 2.066 & 10*     & $3.13 \times 10^{59}$ & 0.051   & Uly/KO/NE\\

010222  & 1.477 &         &                     &         & SAX/WFC\\

010921  & 0.45  &         &                     &         & HE/Uly/SAX\\

011121  & 0.36  & 15.04*  & $6.63 \times 10^{57}$ &         & SAX/WFC\\

011211  & 2.14  &         &                     &         & SAX/WFC\\

020405  & 0.69  & 7.52*   & $1.58 \times 10^{58}$ &         & Uly/MO/SAX\\

020813  & 1.25  & 9.02*   & $8.19 \times 10^{58}$ &         & HETE\\

021004  & 2.3   &         &                     &         & HETE\\

021211  & 1.01  &         &                     &         & HETE\\

030226  & 1.98  & 0.48*   & $1.35 \times 10^{58}$ &         & HETE\\

030323  & 3.37  & 0.0048* & $4.91 \times 10^{56}$ &         & HETE\\

030328  & 1.52  & 2.93*   & $4.31 \times 10^{58}$ &         & HETE\\

030329  & 0.168 & 0.0009* & $7.03 \times 10^{52}$ &         & HETE
\enddata
\tablenotetext{a}{Compiled from S. Barthelmy's IPN redshifts and fluxes
  (http://gcn.gsfc.nasa.gov/gcn/) and J.C. Greiner's catalogue on GRBs
  localized with WFC (BeppoSax), BATSE/RXTE or ASM/RXTE, IPN, HETE-II
  or INTEGRAL (http://www.mpe.mpg.de/~jcg/grbgeb.html)}
\tablenotetext{b}{in cm$^{-2}$s$^{-1}$}
\tablenotetext{c}{Photon luminosities in s$^{-1}$ derived from the measured redshifts and observed gamma-ray fluxes for the cosmological model described in \S2}
\tablenotetext{d}{Opening angles $\theta_j$ in the GRB-emissions refer to the 
  sample listed in Table I of Frail et al.(2001).}
\tablenotetext{*}{Extrapolated to the BATSE energy range 50 - 300 keV using the formula given in 
  Appendix B of Sethi et Bhargavi (2001)}
\end{deluxetable}
\begin{deluxetable}{lllr}
\tablecaption{Model predictions\tablenotemark{a}
              ~versus observations GRB-SNe}
\tablewidth{0pt}
\tablehead{Quantity & Units & Expression & Observation}
\startdata
$E_{gw}$   & erg & $4\times10^{53}~\eta_{0.1}M_{H,7}$ & ~\\
$f_{gw}$   & Hz  & 500~$\eta_{0.1}M_7^{-1}$       & \\
$\Omega_B$ & 1   & $6\times 10^{-9}$@250Hz        & \\
$E_{SN}$   & erg & $2\times 10^{51}~\beta_{0.1}\eta_{0.1}^2M_{H,7}$ & 
                               $2\times10^{51}$erg\tablenotemark{b}\\
$E_\gamma$ & erg & $2\times10^{50}~\epsilon_{0.15}\eta_{0.1}^{8/3}M_{H,7}$ & 
                               $3\times10^{50}$erg\tablenotemark{c}\\
$E_{\gamma\rightarrow X}$\tablenotemark{d} 
                 & erg & $4\times10^{52}~\bar{\epsilon}\eta_{0.1}^{2}M_{H,7}$ & 
                               $>4.4\times10^{51}$erg\tablenotemark{e}\\
$T_s$      & s   & 90~$\eta_{0.1}^{-8/3}M_{H,7}\mu_{0.03}^{-1/2}$ &  
                               $T_{90}$ of tens of s\tablenotemark{f} \\
Event rate & yr$^{-1}$ &         & 1 within $D=100$Mpc\tablenotemark{g}
\enddata
\tabletypesize{\scriptsize}
\tablenotetext{a}{Based on a critical ratio 
  ${\cal E}_B/{\cal E}_k\simeq1/15$ of poloidal magnetic field energy-to-kinetic
  energy in the torus with ratios $b/R<0.3260$ of minor-to-major radius}
\tablenotetext{b}{SN1998bw with aspherical geometry, estimated by H\"oflich et al.(1999)}
\tablenotetext{c}{True energy in gamma-rays produced along open magnetic flux-tubes; Frail et al.(2001)}
\tablenotetext{d}{Continuum gamma-ray emission produced by torus winds with
undetermined efficiency $\bar{\epsilon}$ as energy input to X-ray line-emissions} 
\tablenotetext{e}{Ghisellini et al.(2002)}
\tablenotetext{f}{Kouveliotou et al.(1993)}
\tablenotetext{g}{Local estimate based on Frail et al.(2001) and 
                  van Putten \& Regimbau et al.(2003)}
\end{deluxetable}

\newpage
\centerline{\bf Figure captions}

{\bf FIGURE 1.}
{Cartoon of the proposed model for GRB-supernovae from rotating black holes
(not to scale): core-collapse in an evolved massive star produces an
active MeV-nucleus consisting of a rotating black hole \citep{woo93,bro00} 
surrounded by a torus which may be magnetized with the magnetic field
of the progenitor star \citep{pac98}. The torus assumes a state of suspended
accretion, wherein it catalyzes black hole-spin
energy at an efficiency given by the ratio $\eta=\Omega_T/\Omega_H$ of the
angular velocity $\Omega_T$ of the torus and $\Omega_H$ of the black hole.
Because the nucleus is relativistically compact, the torus radiates this 
input predominantly into gravitational radiation, and, to a lesser degree,
into magnetic winds and MeV-neutrino emissions. 
A small fraction of about $\theta_H^4$ of black hole-spin energy is released in 
baryon-poor jets along open magnetic flux-tubes along the rotational axis of the 
black hole, where $\theta_H$ denotes the half-opening angle on the event horizon.
This output serves as input to the GRB-afterglow emissions. As these jets punch 
through the remnant stellar envelope \citep{mac99}, the GRB may be delayed by
seconds \citep{woo99}, and thereby appear after the onset of gravitational wave-emissions.
A radiatively driven supernova appears subsequently in response to high-energy
continuum emissions produced by the magnetic torus winds. When the envelope has 
expanded sufficiently to becoming optically thin, X-ray line-emissions may appear
conceivably accompanied by radio emissions.} 

{\bf FIGURE 2.}
{Redshift distributions of the flux-limited sample of 33 GRBs with
individually determined redshifts, the true-but-unseen sample assuming the 
GRB event rate is locked to the star-formation rate (hachured), and the 
sample of detectable GRBs predicted by our model according to a log-normal 
peak-luminosity distribution function (grey).
The continuous line represents the cosmic SFR ($\Lambda-$dominated CDM universe).
The unseen-to-observed event rate for standard sources with anisotropic emissions
is hereby 450. (Reprinted from M.H.P.M. van Putten \& T. Regimbau, 2003)}

{\bf FIGURE 3.}
{A causal spin-connection between the torus and the event horizon
of the black hole arises by virtue of an inner torus of open magnetic
field-lines, equivalent to the connection between 
pulsars and asymptotic infinity when viewed in poloidal cross-section. 
These open magnetic field-lines are endowed with Dirichlet-radiative
boundary conditions. The inner face of the
torus (angular velocity $\Omega_+$) and the black hole (angular velocity $\Omega_H$) 
herein corresponds to a pulsar surrounded by infinity with relative angular velocity
$\Omega_H-\Omega_+$ (Mach's principle). It hereby receives energy and angular
momentum from the black hole, whenever $\Omega_H-\Omega_+>0$. The outer face of the 
torus (angular velocity $\Omega_-$) is equivalent to a pulsar with angular velocity
$\Omega_-$, and it looses energy and angular momentum by the same equivalence.
This spin-connection is established by an approximately uniformly
magnetized torus, represented by two counter-oriented current rings, and, for
rapidly rotating black holes, an equilibrium magnetic moment of the horizon.
In poloidal topology, the magnetic flux-surfaces are illustrated in the approximation
of flat space-time. The dashed line is the separatrix between the flux-surfaces of the
inner and the outer magnetospheres. Moving it by a stretch-fold-cut to infinity leaves
an open magnetic flux-tube subtended by the event horizon of the black hole (not
shown). (Reprinted from M.H.P.M. van 
Putten \& A. Levinson, 2003, ApJ, 584, 937 \copyright2003
University of Chicago Press)}

{\bf FIGURE 4.}
{The observed strain-amplitudes $h_+$ and $h_\times$ are
subject to amplitude modulation by varying
orientation of the torus relative to the line-of-sight, in response to
Lense-Thirring precession. This introduces side-bands about the carrier
frequency $2\Omega_T$, where $\Omega_T$ denotes the angular velocity of
the torus. These side-bands are separated from the carrier frequency
by once (in both $h_+$ and $h_\times$) and twice 
(in $h_+$) the Lense-Thirring frequency. Shown are the strain-amplitudes
for the case of $\iota_0=0,\pi/8,\pi/4,\pi/2$ for a wobbling angle of
30$^o$ and a Lense-Thirring precession frequency of 1/8 of the orbital
frequency. The amplitude corresponds to a source at unit distance,
and the index refers to the number of orbital periods. (Reprinted from
M.H.P.M. van Putten, H.K. Lee, C.H. Lee \& H. Kim (2003).)}

{\bf FIGURE 5.}
{({\em Left}) Shown is the observed flux ${\cal F}(z,\mbox{EdS})/(1+z)^3$ 
in gravitational radiation from GRBs as a function of the SFR [$M_\odot$yr$^{-1}$Mpc$^{-3}$]
in an Einstein-de Sitter universe ($H_0=73$km s$^{-1}$ Mpc$^{-1}$), 
evolved to a per unit of surface area and unit of time at $z=0$. Note the peak at
$z\simeq1$. ({\em Right}) Shown is the spectral flux-density $\Omega_B(f)$
for a cosmological distribution of GRBs from rotating black holes as burst 
sources of gravitational radiation assuming a uniform mass distribution in
the range of $M_H=4-14\times M_\odot$ (top curve) and
$M_H=5-8\times M_\odot$ (lower curve). The results are shown for $\eta=0.1$. 
The extremal value of $\Omega_B(f)$ is in the neighborhood of the
location of maximal sensitivity of LIGO and Virgo.}

{\bf FIGURE 6.}
{GRBs from rotating black holes produce a few tenths of $M_\odot$ in
long duration bursts of gravitational radiation \citep{mvp01a,mvp03b}.
These emissions are parametrized by the black hole
mass $M_H=4-14M_\odot$ and the ratio $\eta\sim0.1-0.15$ of the angular velocities of the 
torus and the black hole. The signal is band limited with relative bandwidth 
$B\simeq10\%$. The dark region shows $h_{\mbox{\tiny char}}B^{1/2}/\sqrt{5}$ of the 
orientation-averaged characteristic dimensionless spectral strain-amplitude 
$h_{\mbox{\tiny char}}$. The source distance is $D=100$Mpc, corresponding to an event 
rate of once per year. The dimensionless strain-noise amplitudes 
$h_{\mbox{\tiny rms}}(f)=\sqrt{fS_h(f)}$ of Initial/Advanced LIGO (lines), 
Initial/Cryogenic VIRGO (dashed; \cite{pun99}) are shown with lines removed, including various 
narrow-band modes of Advanced LIGO (dot-dashed), where $S_h(f)$ is 
the spectral energy-density of the dimensionless strain-noise of the detector. 
Short GRBs from binary black hole-neutron star coalescence may produce similar
energies distributed over a broad bandwidth, ranging from low frequencies during 
inspiral up to 1kHz during the merger phase.}

{\bf FIGURE 7.}
{Shown are the achievable upper bounds on the mass of the black hole in GRB030329 
($z=0.167$, $T_{90}=$25 s) at current LIGO
sensitivity ($S_h^{1/2}=4\times 10^{-22}$Hz$^{-1/2}$),  
assuming a no-detection result in applications of matched filtering (solid), 
in correlating two detectors in narrow-band mode (dot-dashed), and in correlating
two detectors in broad-band mode (dotted). 
The labels refer to signal-to-noise ratios 1,2 and 3. In broad-band mode, 
correlation between the two LIGO Hanford detectors at current
sensitivity would permit placing an upper bound on the black hole mass
in GRB030329 ($D=800$Mpc) of about 150$M_\odot$.}

{\bf FIGURE 8.}
{GRBs from rotating black holes are expected to produce line-emissions in gravitational
radiation which evolve slowly in time, on the timescale of spin-down of the black hole. 
This produces trajectories in the temporal evolution of the spectrum
of the signal. We may search for these trajectories, by performing Fourier transforms
over time-windows of intermediate size, during which the signal is approximately
monochromatic. The results shown illustrate a slowly-evolving
line-emission for a long burst, partitioned in $N=128$ sub-windows of $M=256$ data
points, in the presence of noise with an instantaneous signal-to-noise ratio of 0.15.
The left two windows show the absolute values of the Fourier coefficients, 
obtained from two simulated detectors with uncorrelated noise. 
The trajectory of a simulated slowly-evolving emission-line
becomes apparent in the correlation between these two spectra (right window).
The frequency scales with Fourier index $i$ according to $f=(i-1)/\tau$
($i=1,\cdots M/2+1$), where $\tau$ denotes the time-period of the sub-window.}

\newpage
\begin{figure}
\plotone{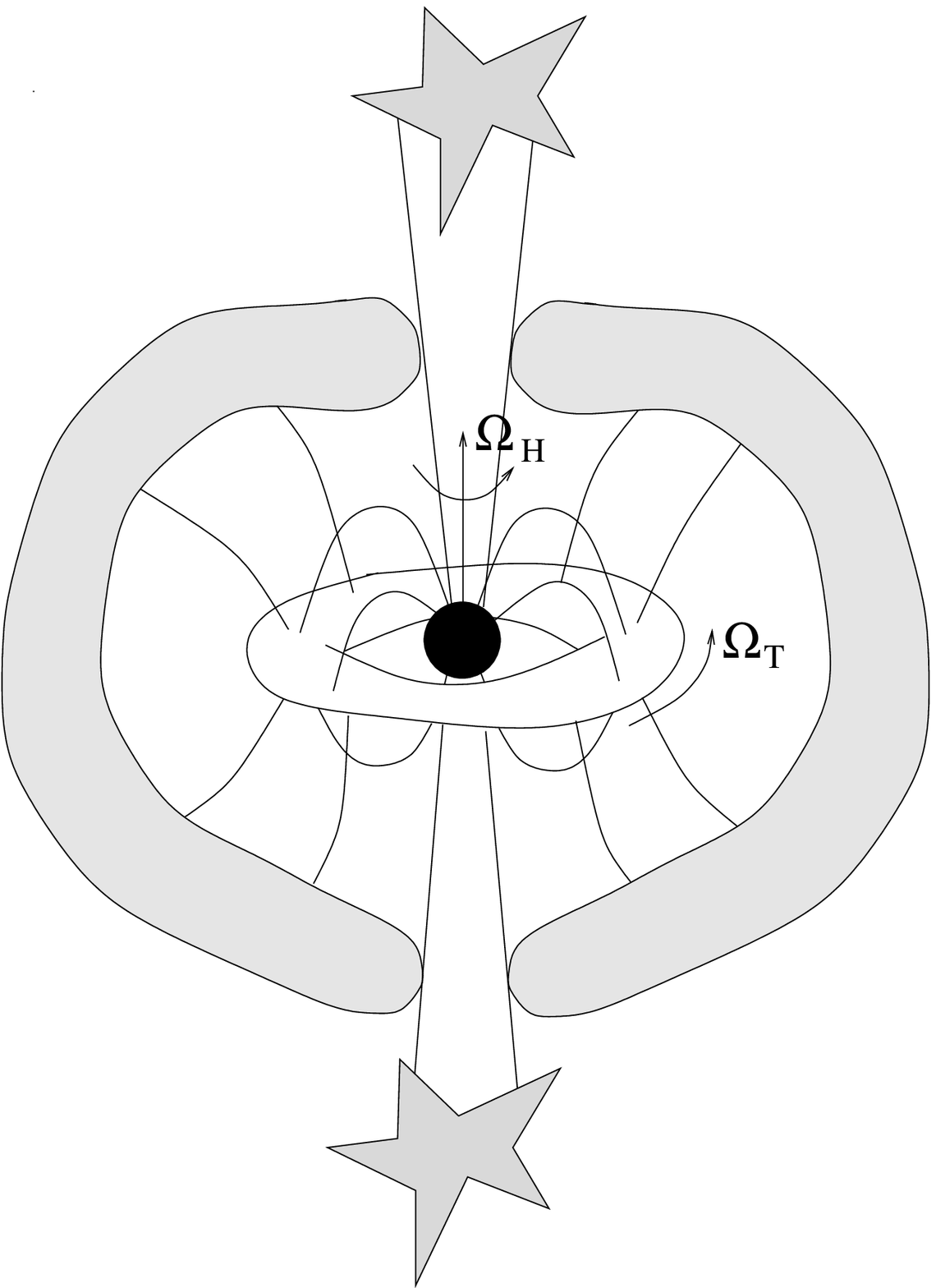}
\caption{}
\end{figure}
\begin{figure}
\plotone{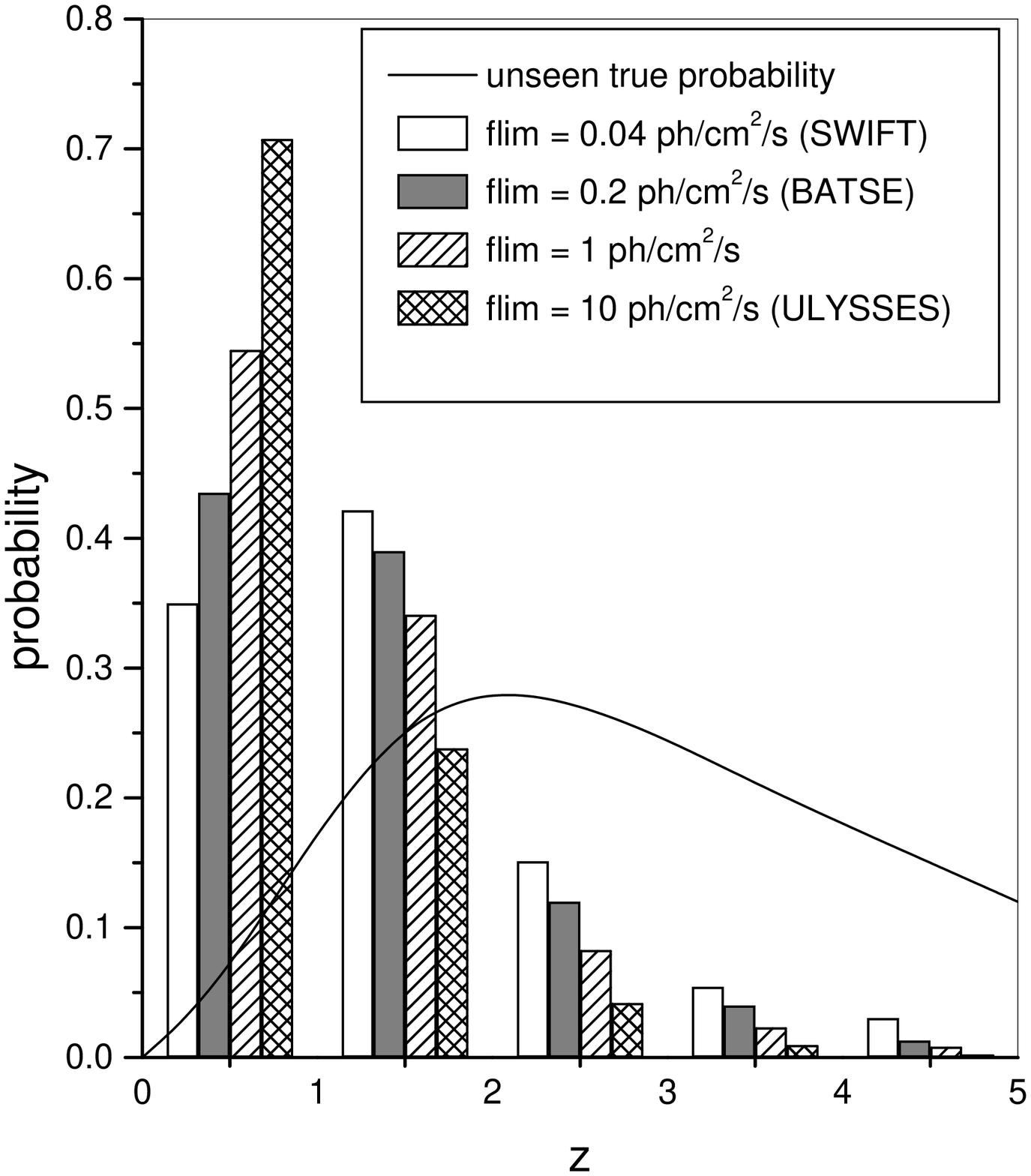}
\caption{}
\end{figure}
\begin{figure}
\plotone{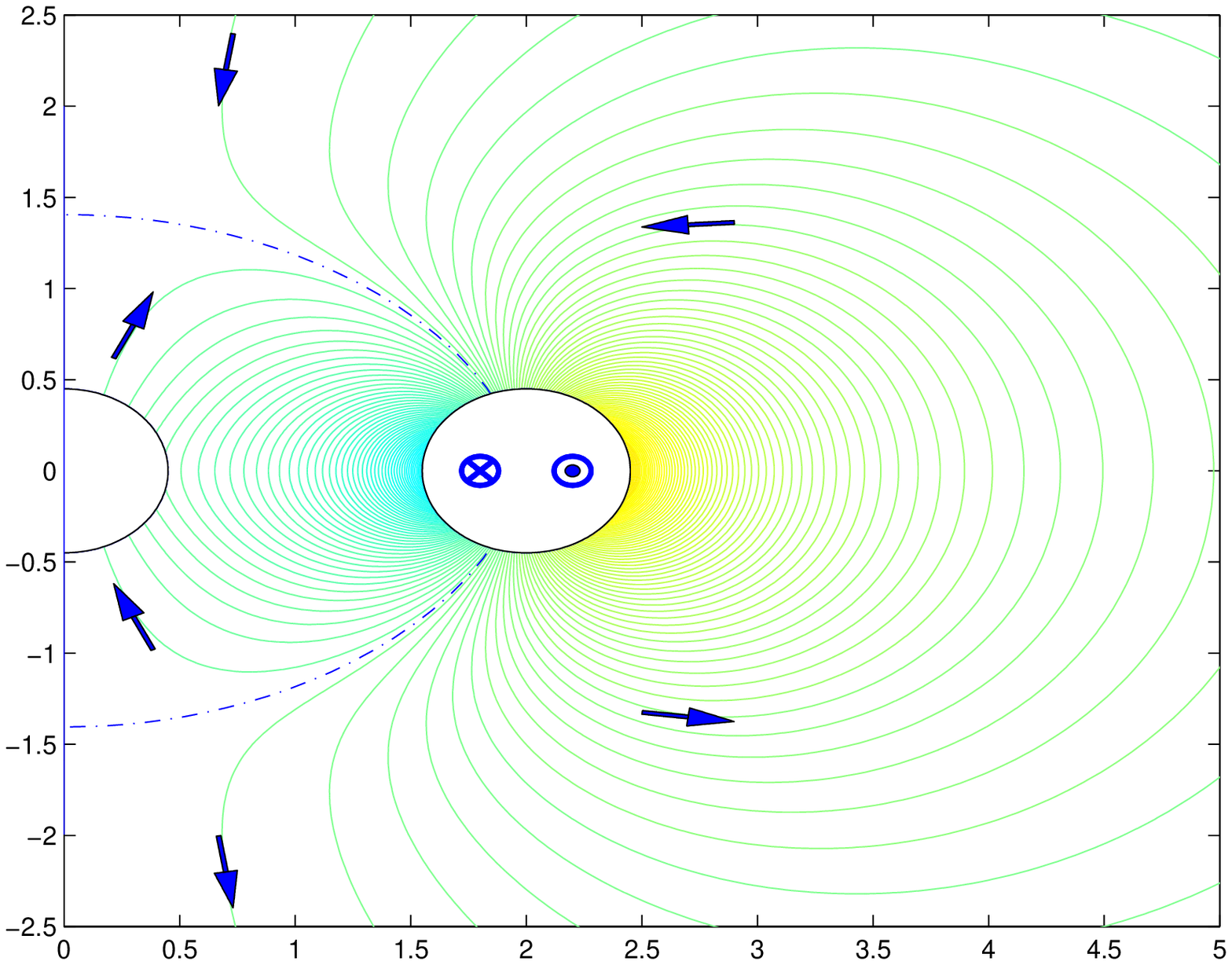}
\caption{}
\end{figure}
\begin{figure}
\plotone{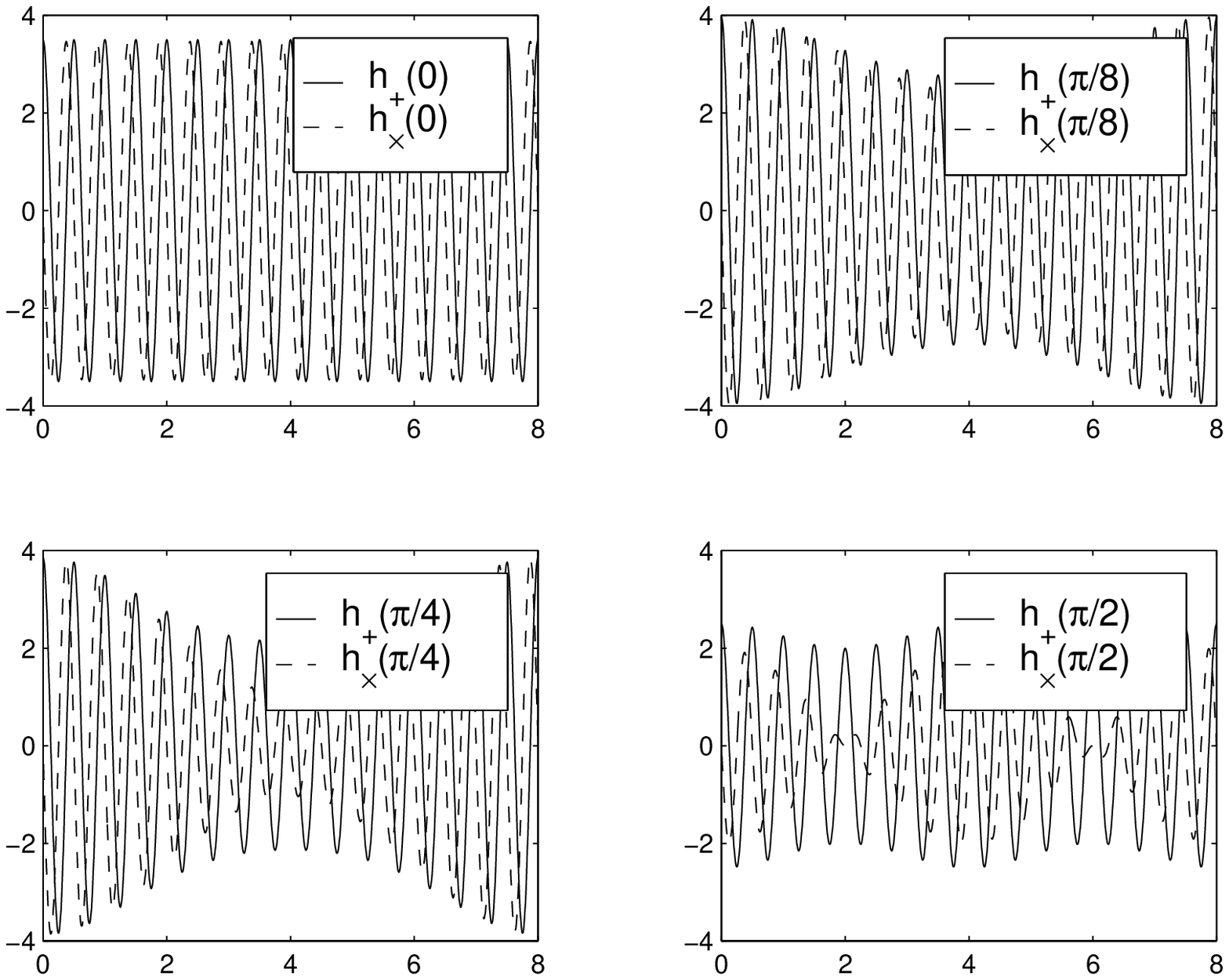}
\caption{}
\end{figure}
\begin{figure}
\plotone{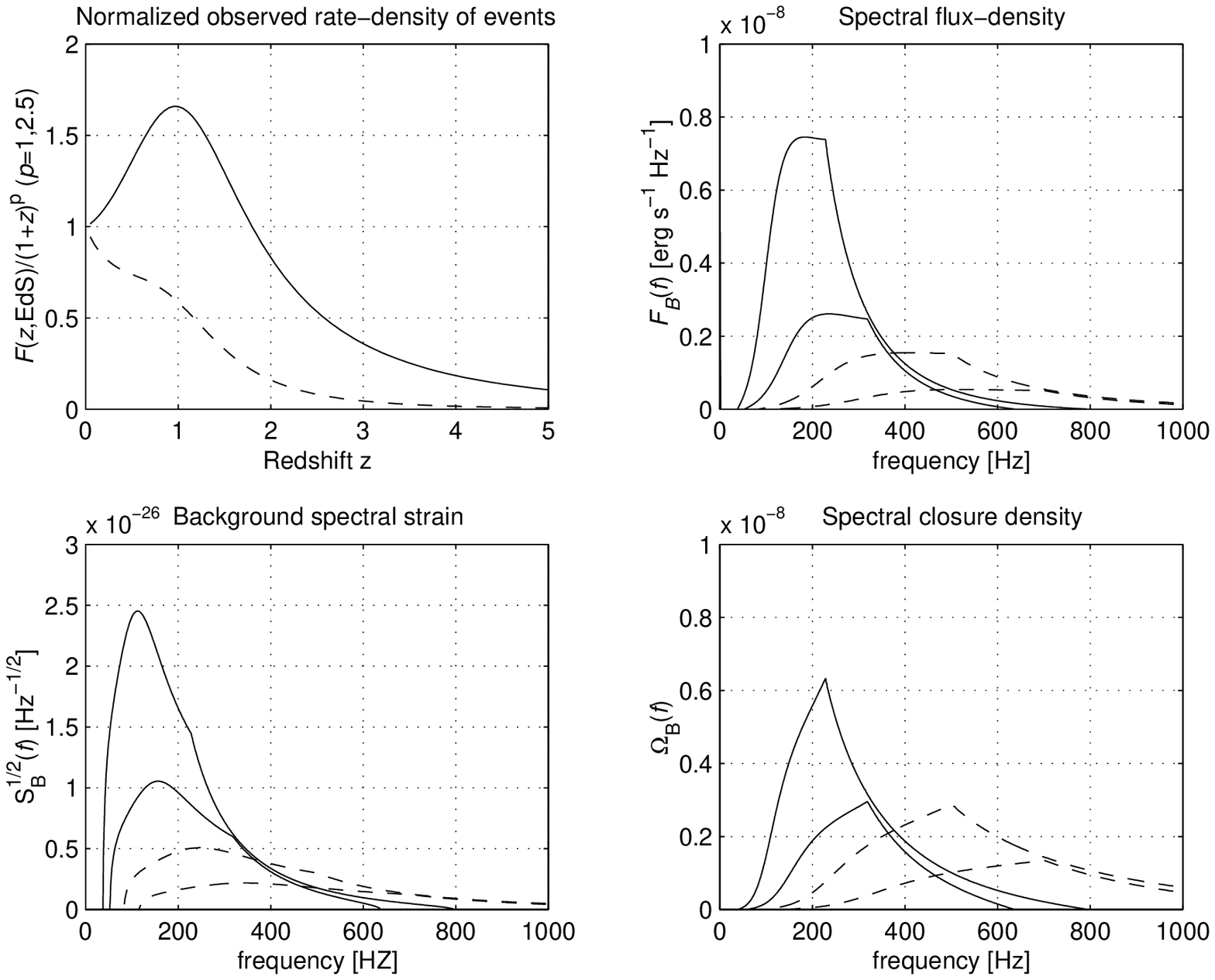}
\vskip0.1in
\caption{}
\end{figure}
\begin{figure}
\plotone{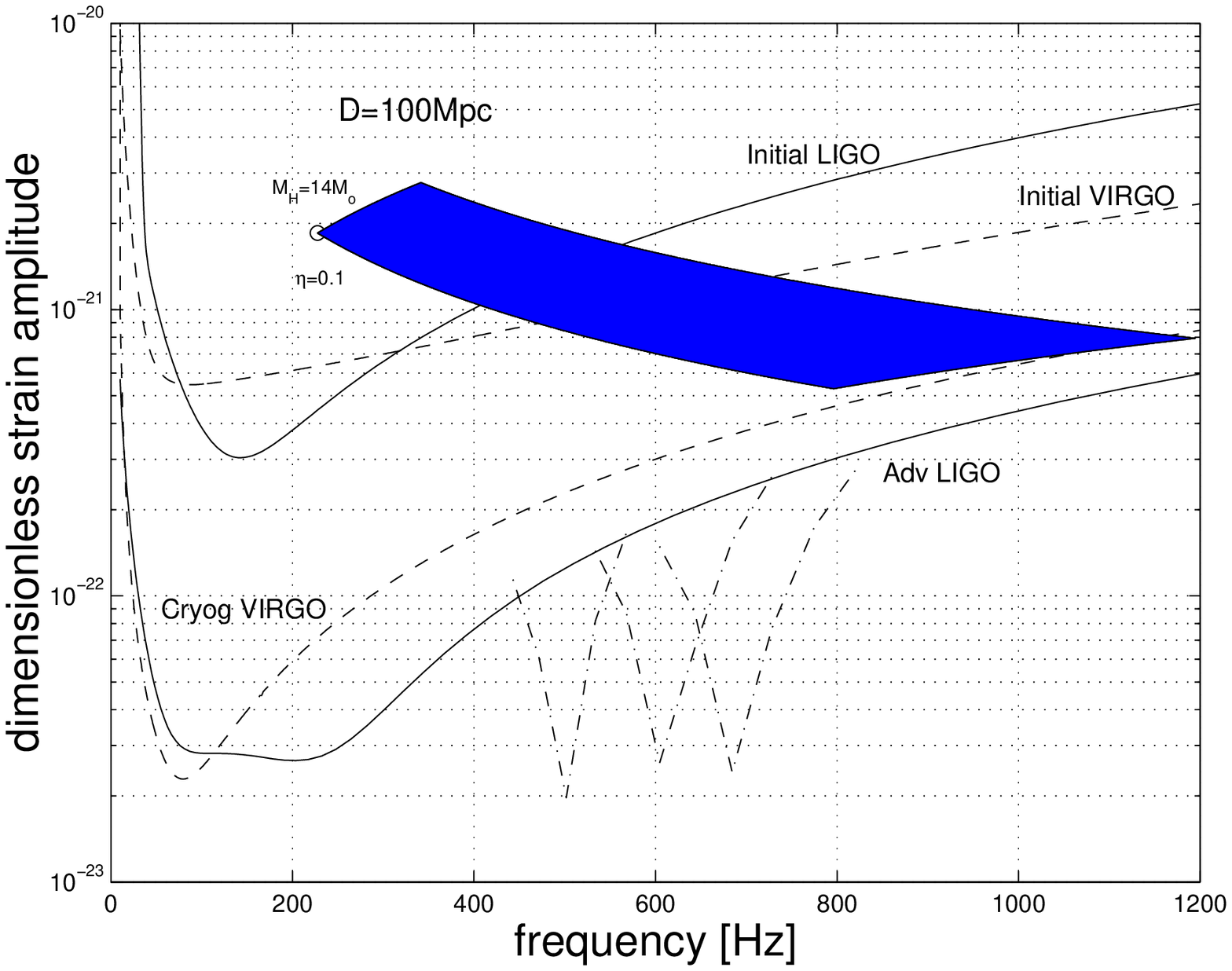}
\caption{}
\end{figure}
\begin{figure}
\plotone{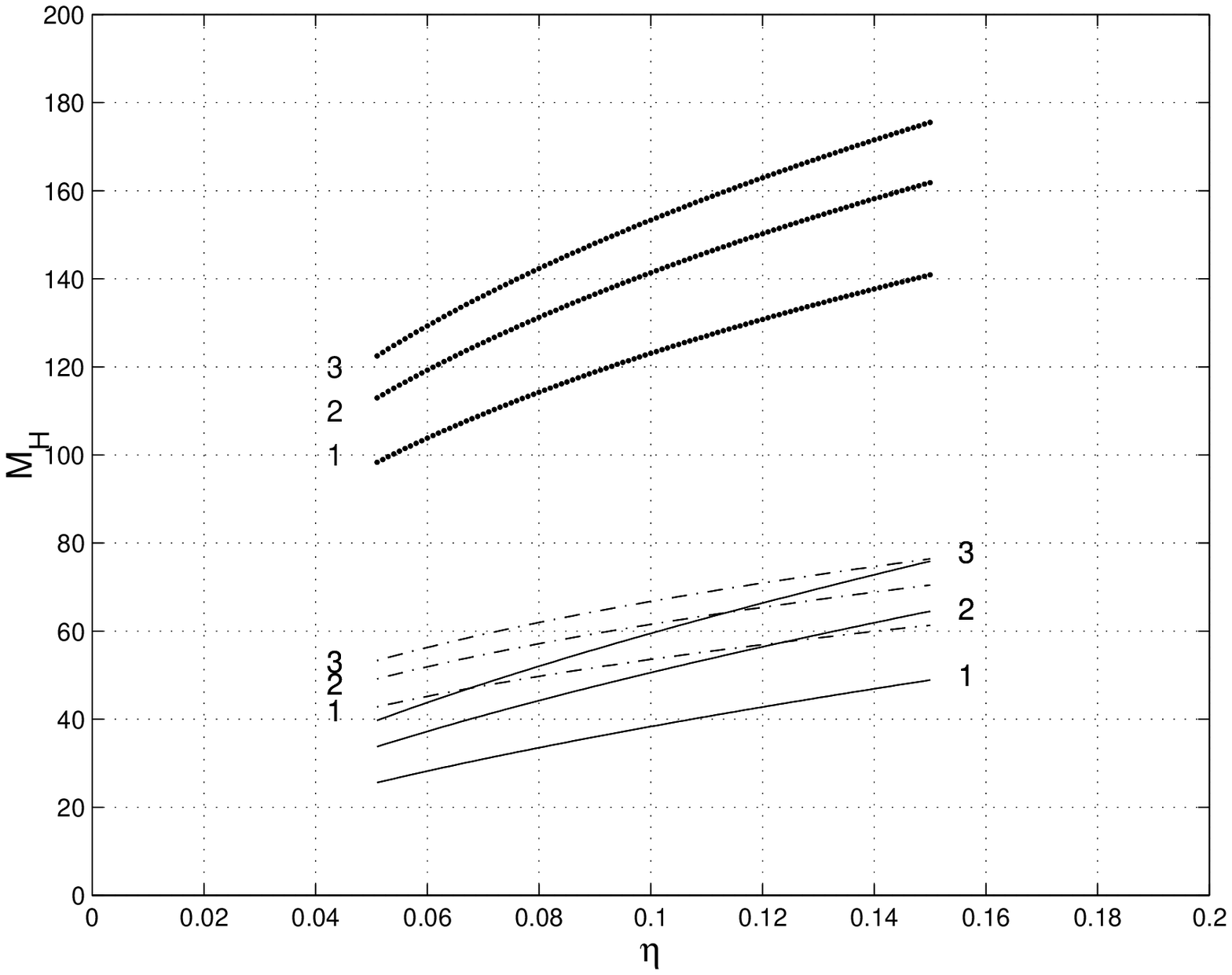}
\caption{}
\end{figure}
\begin{figure}
\plotone{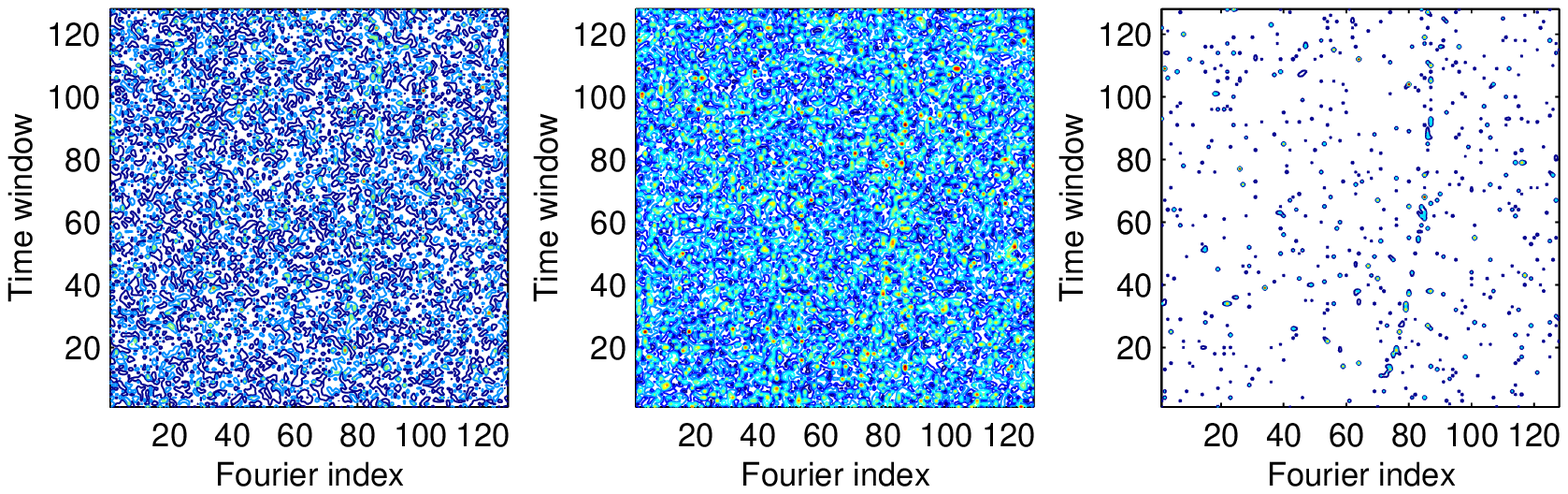}
\caption{}
\end{figure}
\end{document}